\documentclass[a4paper,fleqn,usenatbib,useAMS]{mnras}
\usepackage{graphics}
\usepackage{graphicx}
\usepackage{amsmath}    
\usepackage{amssymb}    
\usepackage{multicol}        
\usepackage{bm}     
\usepackage{pdflscape}  
\usepackage{xcolor}
\usepackage{ae,aecompl}
\usepackage[english]{babel}
\usepackage{times}
\usepackage{etoolbox}
\def\be{\begin{equation}}
\def\ee{\end{equation}}
\usepackage[utf8]{inputenc}
\usepackage{lmodern}
\definecolor{vale}{rgb}{1.,.5, 0.0}

\newif\ifAMStwofonts
\AMStwofontstrue
\graphicspath{{./fig/}}

\title[Polarimetry microlensing of stellar atmosphere]{Measuring the stellar atmosphere parameters using follow-up polarimetry microlensing observations}
\author[Khalouei et al.]{
Elahe Khalouei,$^{1}$\thanks{E-mail: e.khalouei1991@gmail.com}
Sedighe Sajadian,$^{2,3}$
Sohrab Rahvar$^{1}$\\
$^{1}$Department~of~Physics,~Sharif~University~of~Technology,~P. O. Box 11365-9161,~Tehran,~Iran\\
$^{2}$Department~of~Physics,~Isfahan~University~of~Technology,~Isfahan~84156-83111,~Iran\\
$^{3}$Department~of~Physics,~Chungbuk~National~University,~Cheongju~28644,~Republic~of~Korea}
 
\begin{document}
\label{firstpage}
\maketitle	
\begin{abstract}	
We present an analysis of the potential follow-up polarimetry microlensing observation to study the stellar atmospheres of the distant stars. First, we produce synthetic microlensing events using the Galactic model, stellar population, and interstellar dust toward the Galactic Bulge. We simulate the polarization microlensing light curves and pass them through the instrument specifications of FOcal Reducer and low dispersion Spectrograph (FORS2) polarimeter at Very Large Telescope (VLT), and then analyze them. We find that the accuracy of the VLT telescope lets us constrain the atmosphere of cool RGB stars. Assuming detection of about $3000$ microlensing events per year by the OGLE telescope, we expect to detect almost $20,~10,~8, $ and $5$ of polarization microlensig events for the four different criteria of being three consecutive polarimetry data points above the baseline with $1\sigma$, $2\sigma$, $3\sigma$, and $4\sigma$, respectively in the polarimetry light curves. We generalize the covariance matrix formulation and present the combination of polarimetry and photometry information that leads us to measure the scattering optical depth of the atmosphere and the inner radius of the stellar envelope of red giant stars. These two parameters could determine the dust opacity of the atmosphere of cool RGB source stars and the radius where dust can be formed.
\end{abstract}
 

\section{Introduction}\label{one}

Gravitational microlensing is being used as a strong tool for studying the distribution of matter and dark matter in the Milky Way galaxy \citep{Paczynski1986}. It is also being used for searching the exoplanets as well as studying the atmosphere of stars \citep[see, e.g.,][]{gaudi2012,rahvar2015}. This effect is based on the bending of the light path from a background source star with the gravitational field of foreground massive objects that produces two distorted images. This effect also magnifies the luminosity of a source star during the gravitational lensing \citep{Einstein1936}. The magnification factor, i.e. the ratio of the magnified to the baseline luminosity of a point-like source star is given by \citep[see, e.g.,][]{Paczynski1986}:
\begin{equation}
A(t)=\frac{u^{2}(t)+2}{u(t)\sqrt{u^{2}(t)+4}},
\end{equation}
where $u(t)=\sqrt{u_{0}^{2}+\left[\frac{t-t_{0}}{t_{\rm{E}} }\right]^{2}}$ is the projected distance between the source and lens on the lens plane normalized to the Einstein radius where $u_{0}$ is the minimum impact parameter. $t_{0}$ is the time of the closest approach and  $t_{\rm{E}}$ is the Einstein crossing time. Here, the Einstein radius and Einstein crossing time are given by:
\begin{eqnarray}
R_{\rm{E}}=\sqrt{\frac{4~G~M_{l}}{c^{2}}~D_{l}~(1-\frac{D_{l}}{D_{s}})} ,\qquad
t_{\rm{E}}={R_{\rm{E}}}/{v_{T}},
\end{eqnarray}
where $G$ is the gravitational constant, $M_{l}$ is the lens mass, $c$ is the speed of the light, $v_{T}$ is the relative lens-source transverse velocity, $D_{\rm s}$ and $D_{\rm l}$ are the distances of the source and lens from the observer, respectively. This relation represents that there is a degeneracy between the physical parameters of the lens and source stars in the microlensing observation \citep[see, e.g.,][]{skowron2012, Kains2013}.

Other observations such as astrometry, parallax, Xallarap, finite-size effects, and polarimetry observations can break this degeneracy \citep{poindexter,rdominik}.  In this work, we study the possibility of polarimetry observations of on-going microlensing events. With these observations, we can break the degeneracy and obtain extra information about the parameters of the atmosphere of the source stars.

In the microlensing observations towards the Galactic bulge, most source stars are in the bulge. While for this direction the interstellar dust prevent observations of part of source stars, however along some 
line of sights the stars of the bulge is observable. The scattering of photons in the stellar atmospheres makes the starlight being locally polarized \citep{chandrasekhar60} and due to the circular symmetry of the surface of the star, the overall polarization is zero. In gravitational microlensing, the circular symmetry of the source disc is broken which results in a net polarization signal. Polarization in high-magnification and caustic-crossing of microlensing of supernovae was first theoretically evaluated by \cite{schneider1987}. Then, \citet{simmons1995a,simmons1995b,Bogdanov1996} studied the polarization signals for microlensing events numerically. \citet{agol1996} investigated the polarization in binary microlensing events and discussed the benefits of polarimetry observations to resolve the microlensing degeneracies. Also, \citet{yoshida2006} obtained analytical expressions of polarization in the microlensing events by considering the finite source size. All of the mentioned researches assumed that the microlensing sources are early-type stars. It has been also considered the red clump giant (RCG) as a source in microlensing events and developed a formalism to evaluate the polarization in microlensing of this type of stars \citep{simmons2002, Ignace2006, Ignace2008}. \citet{Ingrosso2012, Ingrosso2015} extended these works and studied the polarization in microlensing of late-type main-sequence and RCG sources in addition to early-type stars. They also evaluated the expected polarization signals of the reported OGLE-III microlensing events \citep{Wyrzykowski2015b}. 
As noted by \citet{Ing}, the polarization observations could be an efficient tool to clarify the binary microlensing model due to specific features of polarization curves. In this work, we investigate 
 \begin{itemize}
 \item
 How polarimetry observation of microlensing events is feasible to  (i)  
 measure the stellar atmospheric parameters and (ii) reduce the relative errors in $t_{E}$, $u_{0}$, and $\rho_{*}$ microlensing parameters which are determined by photometry observations.
 \item
 The type of source stars that could be efficient in detection of the polarization signals in microlensing events. \item 
 The expected number of microlensing events with detectable polarization signals per year by potential follow-up polarimetry observations. 
\end{itemize}  
  In this regard, we perform a realistic Monte-Carlo simulation toward the Galactic bulge with the microlensing surveys such as OGLE\footnote{\url{http://ogle.astrouw.edu.pl/ogle4/ews/ews.html}}, MOA\footnote{\url{http://www.phys.canterbury.ac.nz/moa/}}, and KMNet\footnote{\url{http://kmtnet.kasi.re.kr/kmtnet-eng/}}. Then, we simulate synthetic polarization observations of these events with FORS2 polarimeter at VLT.  For each simulated event, we fit the synthetic microlensing light curves with the theoretical model. Moreover, we calculate the covariance and fisher matrix to evaluate the uncertainties of the physical parameters of the source atmosphere.

The framework of the paper is organized as follows. In section \ref{two}, we revisit the polarization formalism in microlensing events for different types of source stars, i.e., main sequence and Red Giant Clump (RGC) stars. In section \ref{three}, we explain the details of our Monte-Carlo simulation. Section \ref{five} is the results of the simulation include (i) the statistics of microlensing events with detectable polarization signals and (ii) the error analysis of the stellar physical parameters. Section \ref{summary} summarizes this work.
\begin{figure*}
\centering
\includegraphics[angle=0,width=0.49\textwidth,clip=]{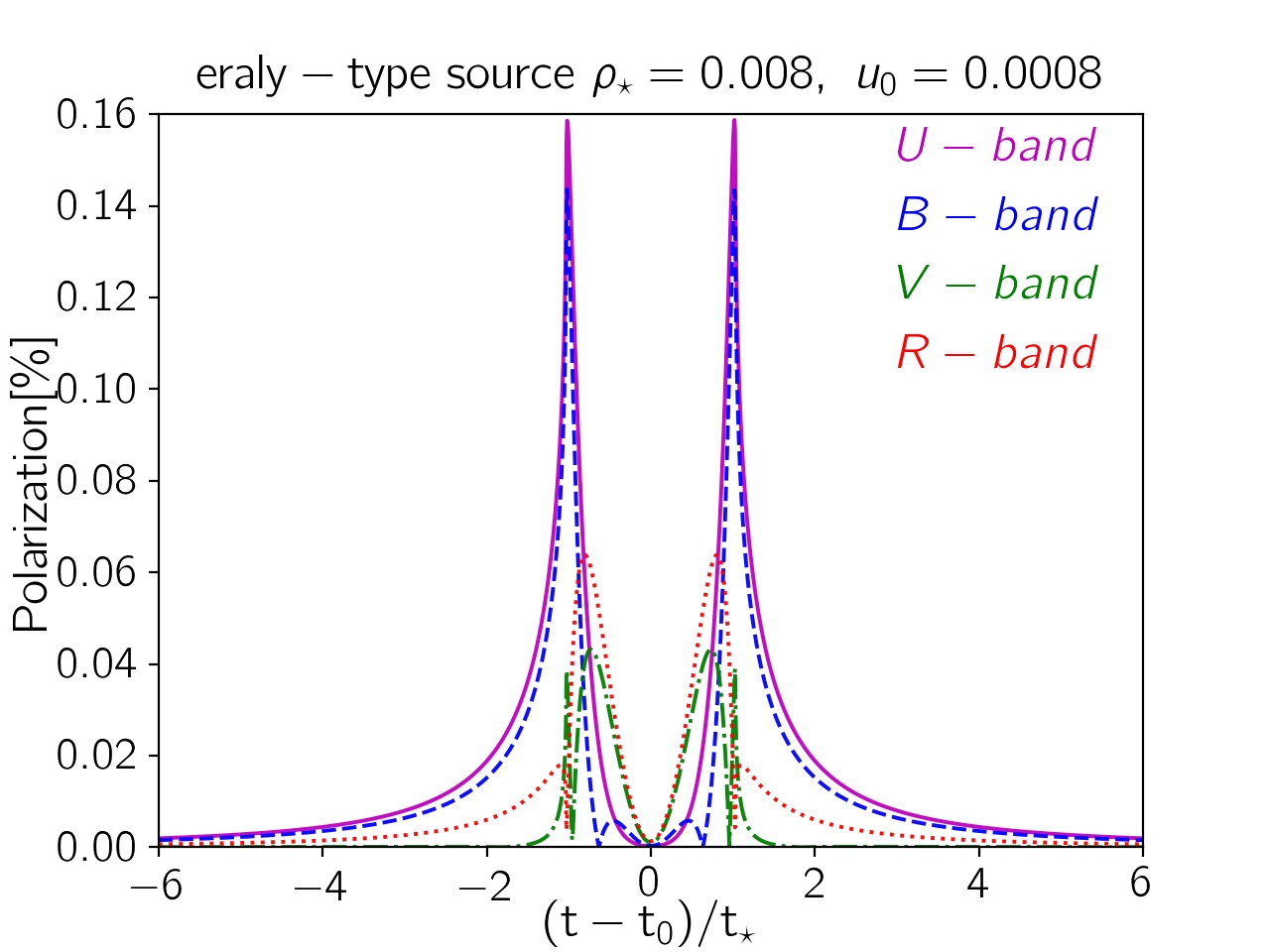}
\includegraphics[angle=0,width=0.49\textwidth,clip=]{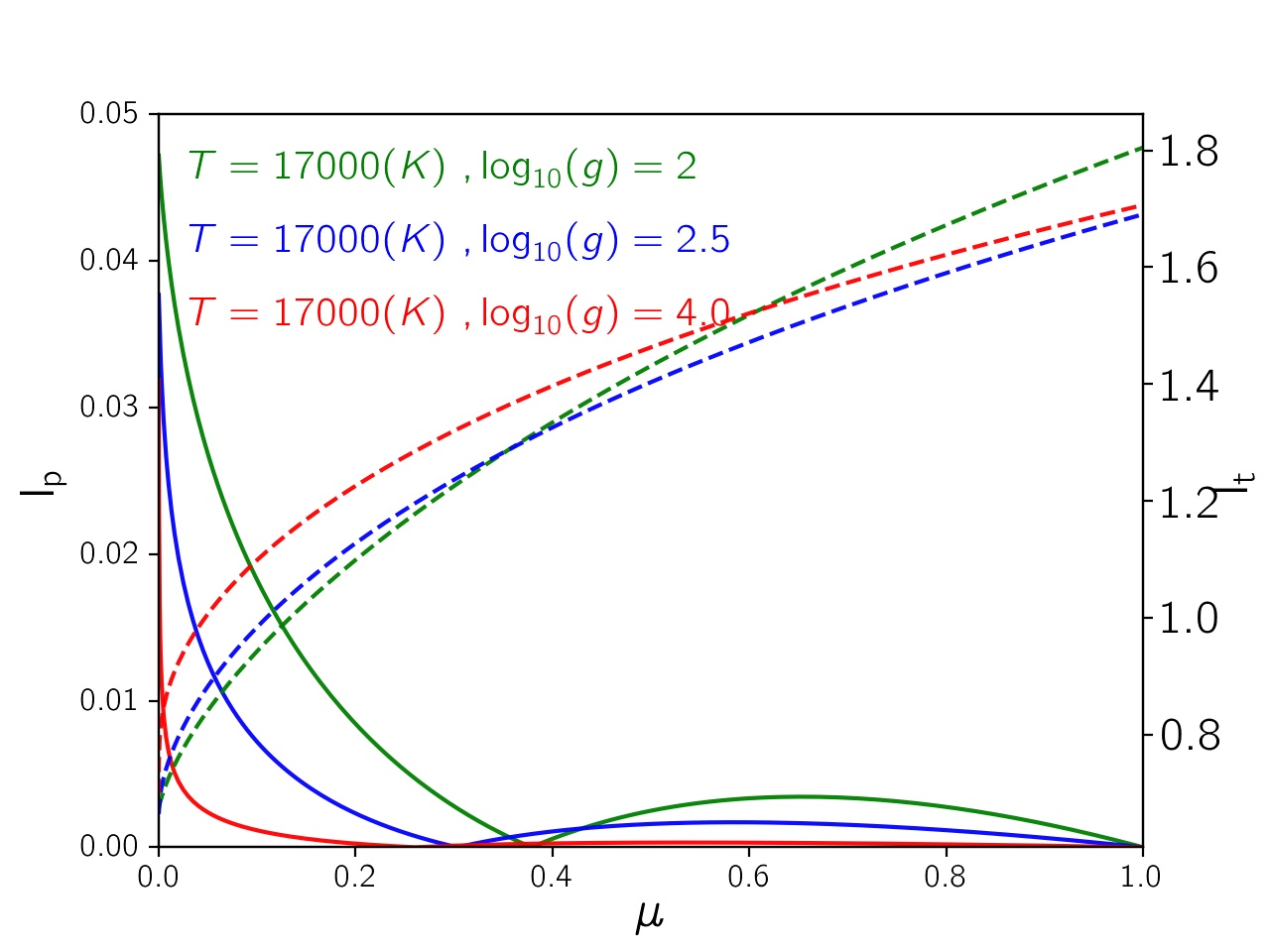}
\caption{Left panel: Polarization microlensing light curves by an early-type source star with four filters of $UBVR$ in the standard photometry system. The x-axis is the time normalized to $t\star$.  The right panel: Polarization (solid line) and light intensity (dashed line) in terms of $\mu$ for the early-type source stars with different amounts of surface gravity. The polarization intensity is  defined as $I_{p} = I_{r} - I_{l}$ where $I_{r}$ and $I_{l}$ are the radial and tangential intensity. The sign of $I_{p}$ could change from the center to the edge of the stellar disk.}\label{hotstar}
\end{figure*}

\section{Polarization in microlensing events}\label{two}
To describe the polarized light, we use the formalism of the Stokes parameters \citep{Tinbergen1996} in gravitational microlensing \citep{simmons2002}, 
\begin{equation}
\left ( \begin{array}{c}
S_{Q}\\
S_{U}\\
S_{I}
\end{array}\right)=
\rho_{\star}^{2}\int_{0}^{1} \rho d\rho \int_{-\pi}^{\pi } d\phi
A(\mu) \left ( \begin{array}{c}
-I_{p}(\mu)\cos(2 \phi)\\
-I_{p}(\mu) \sin(2 \phi)\\
I_{t}(\mu)
\end{array}\right),
\end{equation}
here $S_{\rm I}$, $S_{\rm Q}$ and $S_{\rm U}$ are the total and the two components of the linear polarized luminosity, respectively. The integration is taken over source surface with the projected radius $\rho_{\star}$ normalized to the Einstein radius, $\rho$ is radial component over the surface of the star which is normalized to $\rho_{\star}$, $\mu=\sqrt{1-\rho^{2}}$ and $\phi$ is the azimuthal angle over the source disc. $I_{t}(\mu)$ and $I_{p}(\mu)$ are the total and polarized starlight intensities. The polarization signal and the angle of polarization are given by the following expressions \citep{chandrasekhar60}
\begin{equation}
P=\frac{\sqrt{S_{Q}^{2}+S_{U}^{2}}}{S_{I}},
\end{equation}
\begin{equation}
\theta_{p}=\frac{1}{2} \tan^{-1}\frac{S_{U}}{S_{Q}}.
\end{equation}
In the following subsections, we study the polarization curves for two stellar types of (i) main sequence and (ii) red clump stars. For the main sequence stars, we have three categories of (i) hot stars with $T>15000~$K where the polarization is dominated by the free electrons, (ii) the cool stars with $T<8000~$K where the polarization is sourced by molecules in the atmosphere and (iii) the stars with intermediate temperature $8000<T<15000~$K where the polarization is small. 

\subsection{Main Sequence stars}

Due to the high stellar temperature of hot stars with $T>15000~$K, there are free electrons in their atmospheres and Thomson scattering is the dominant effect in producing the polarized light. \citet{chandrasekhar60} first evaluated the Stokes intensity parameters for this type of star. His calculation was based on pure scattering without taking into account the absorption effect. Regarding the analytical solution of Chandrasekhar, \citet{schneider1987} and others fitted two linear functions as the total and polarized Stokes parameters versus $(1-\mu)$:
\begin{eqnarray}\label{eqpol}
I_{t}(\mu)&=&I_{0}(1-c_{1}(1-\mu)),\nonumber\\
I_{p}(\mu)&=&I_{0}c_{2}(1-\mu),
\end{eqnarray}
where $I_{0}$ is the intensity at the center of the stellar disc. The total intensity of $I_{t}(\mu)$ decreases from center to the limb and $c_1$ and $c_2$ are the limb-darkening coefficients. Generally, the limb-darkening coefficients can be evaluated by solving the transfer equations in real atmospheric models. For instance, based on ATLAS and PHOENIX models, \citet{claret2011,claret2018,claret2019} evaluated the limb-darkening and gravity-darkening coefficients for different stellar types in many photometry systems. Another theoretical investigation is done by \citet{Harrington15} who estimated the Stokes intensities versus monochromatic wavelengths for the hot stars with $T>15000~$K 
based on the NLTE stellar atmosphere model of TLUSTY code \citep{tlusty2,tlusty1}.  

For cool stars, Rayleigh scattering by molecular, atoms, and dust with the main contribution and Thomson scattering with the minor contribution makes starlight be polarized. The analytical equation of the polarization degree for a Sun-like star was studied by \citet{stenflo2005}. For stars cooler than the Sun, we expect more polarized intensities compare to the Sun. For cool stars in the temperature range $2500-8000~\rm{K}$, we use the total and polarized Stokes intensities which were evaluated by \citet{Harrington15} based on MARCS atmosphere models \citep{marcs2008}.

\noindent Also, Harrington based on the model of Kurucz \footnote{http://kurucz.harvard.edu/grids.html} has presented the Stokes intensities for the stars with intermediate temperatures (i.e. $8000~\rm{K}$ to $15000~\rm{K}$). The magnitude of the polarization signal depends on the abundance of scattering agents in the upper part of the atmosphere. In cool stars, there is molecular hydrogen, which scatters effectively. In the intermediate temperature stars, there are few free electrons but no hydrogen molecules. The atomic hydrogen does not compete effectively with the $H^{-}$ ion, which is the main source of opacity. Therefore, the polarization signal of stars with intermediate temperatures is less than the other categories of stars. \footnote{https://www.astro.umd.edu/~jph/Stellar$\_$Polarization.html}. According to Harrington outputs, the polarized and limb-darkening coefficients depend on the wavelength, the surface gravity, the stellar temperature, and somewhat the stellar metallicity. For the different stellar types, we first integrate the Stokes intensities over the standard filters $\rm{UBVRI}$ and then fit polynomial functions in terms $(1-\mu)$ for the total and polarized Stokes intensities, as 
\begin{eqnarray}
I_{t}&=&\sum_{i=0}^{8}a_{i} (1-\mu)^{i},\nonumber\\
I_{p}&=&\sum_{i=0}^{8}b_{i} (1-\mu)^{i}.
\end{eqnarray}
In this work, we use the Stokes intensities were provided by  \citet{Harrington15}.
In Table (\ref{tab1}) of Appendix, we summarize the parameters $a_{i}$ and $b_{i}$ for different stellar types with given $T_{\rm{eff}}$  and $\log_{10}[g]$ values in the standard Johnson filters $UBVRI$ which  are represented with the numbers $0$, $1$, $2$, $3$, and $4$, respectively.

In what follows, we use the limb darkening model to plot a typical light curve of the intensity and polarization during the microlensing events. In Figure (\ref{hotstar}), we plot the polarization curves (in $UBVR$ Johnson filters) in terms of time normalized to $t_{\star}=t_{\rm{E}}~\rho_{\star}$, for different surface gravity of an early-type hot star. Here, we adapt $\rho_{\star}=0.008$ while the lens crosses its disc with the impact parameter $u_{0}=0.1~\rho_{\star}$ . We use the codes of \citet{valerio1,valerio2} for calculating the magnification of main sequence source stars. The significant key from this figure is that the polarization signal is higher in the shorter wavelengths. Indeed, in the smaller wavelengths, the radiation is mostly streaming upward (in the direction of the temperature gradient) which results in higher local polarization signals in the U-band at the source edge than in the R-band \citep{Harrington1970, Harrington15}. The maximum polarization signal in microlensing events happen when the lens is crossing the edge of the source, so in the bluer filters, we get more polarization signals.

\noindent In addition to the wavelength, the local Stokes intensities depend on the stellar surface gravity  (see the right panel of Figure (\ref{hotstar})). By increasing the surface gravity, the local polarized Stokes intensity decreases everywhere on the star surface as the atmospheres with higher surface gravities are denser at the depth and the absorption takes place more than the scattering, unless in the outer layer of the atmosphere.

\begin{figure}
	\includegraphics[angle=0,width=0.49\textwidth,clip=]{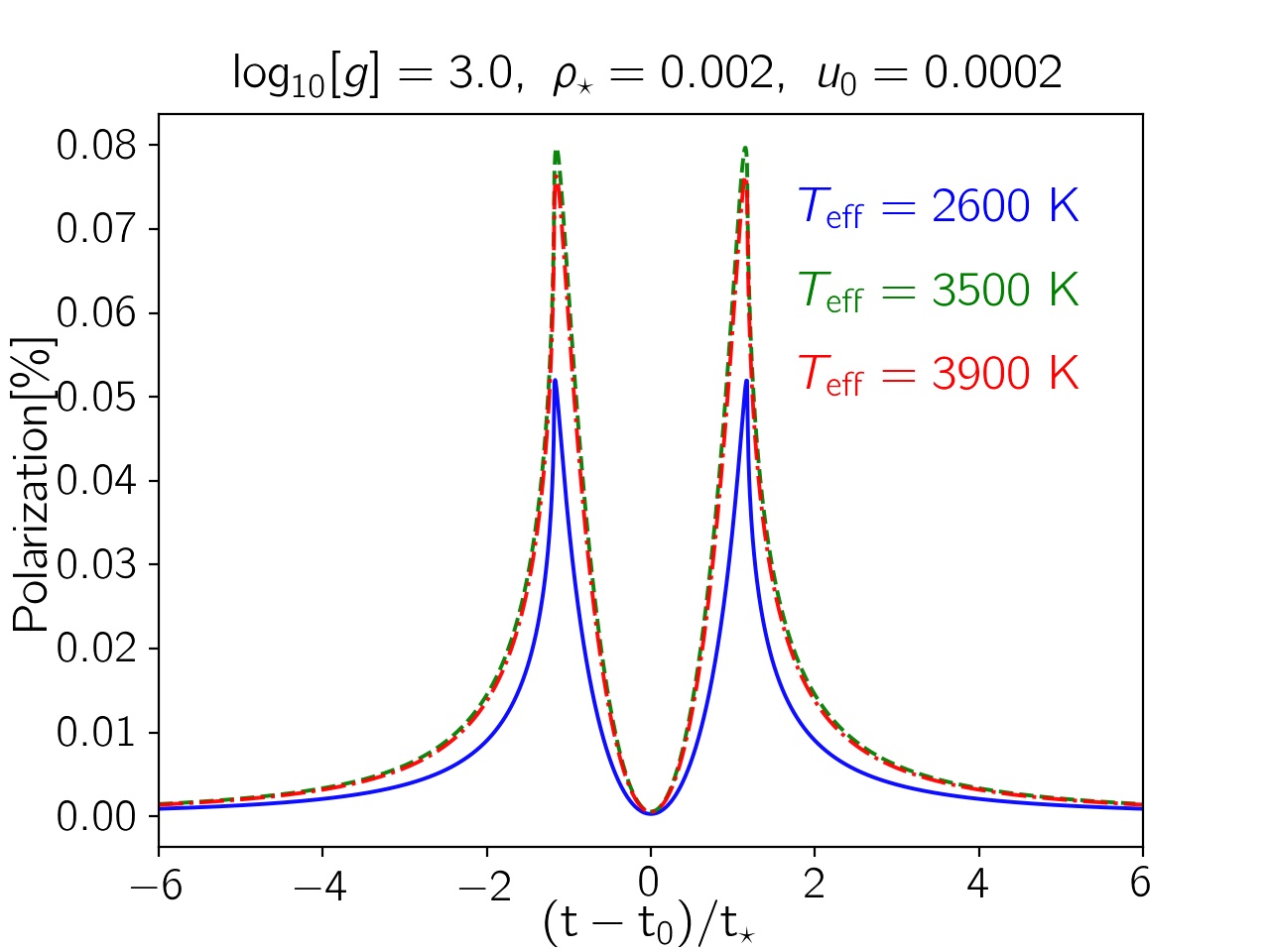}
	\caption{Sample of the polarization curves of cool main-sequence stars with different surface temperatures in a microlensing event.
			}
	\label{coolstar}
\end{figure}

\noindent In Figure (\ref{coolstar}), we plot a sample of polarization curves for a cool main-sequence source star in a microlensing event. The source star for this event is a M-dwarf, cooler than a Sun-like star with a stronger polarization signal.  
 
\begin{figure}
	\centering
	\includegraphics[angle=0,width=0.35\textwidth,clip=]{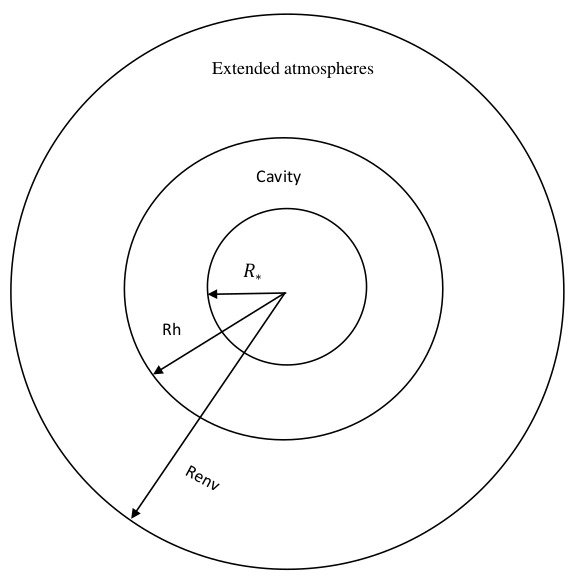}
	\caption{The cross-section of a cool RGB star.  $R_\star$ is the star radius. There is a cavity above the star and a large envelope with the 
	inner and outer radius of $R_h$ and $R_{env}$.}\label{simon}
\end{figure}

\begin{figure*}
	\includegraphics[angle=0,width=0.49\textwidth,clip=]{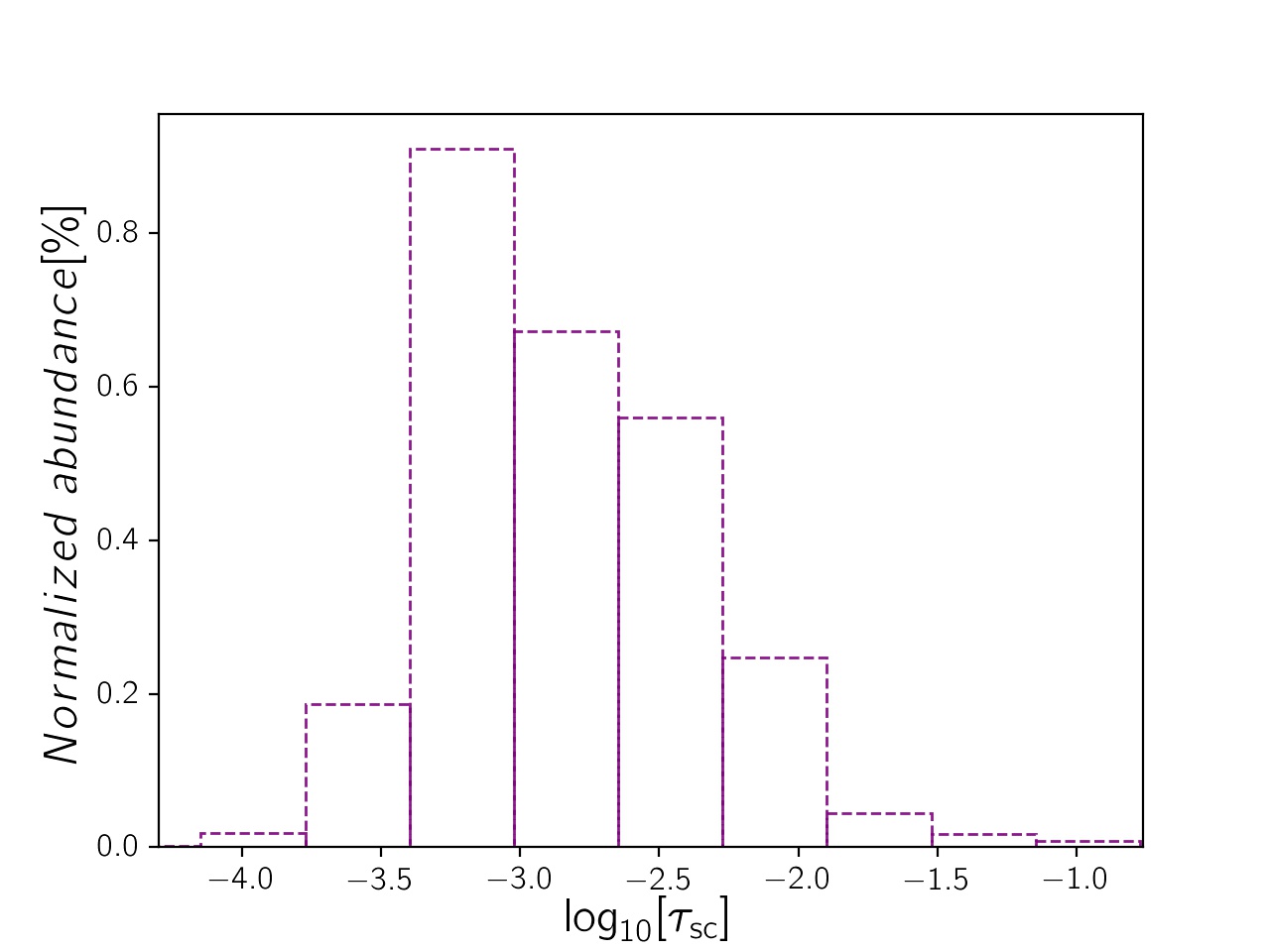}
	\includegraphics[angle=0,width=0.49\textwidth,clip=]{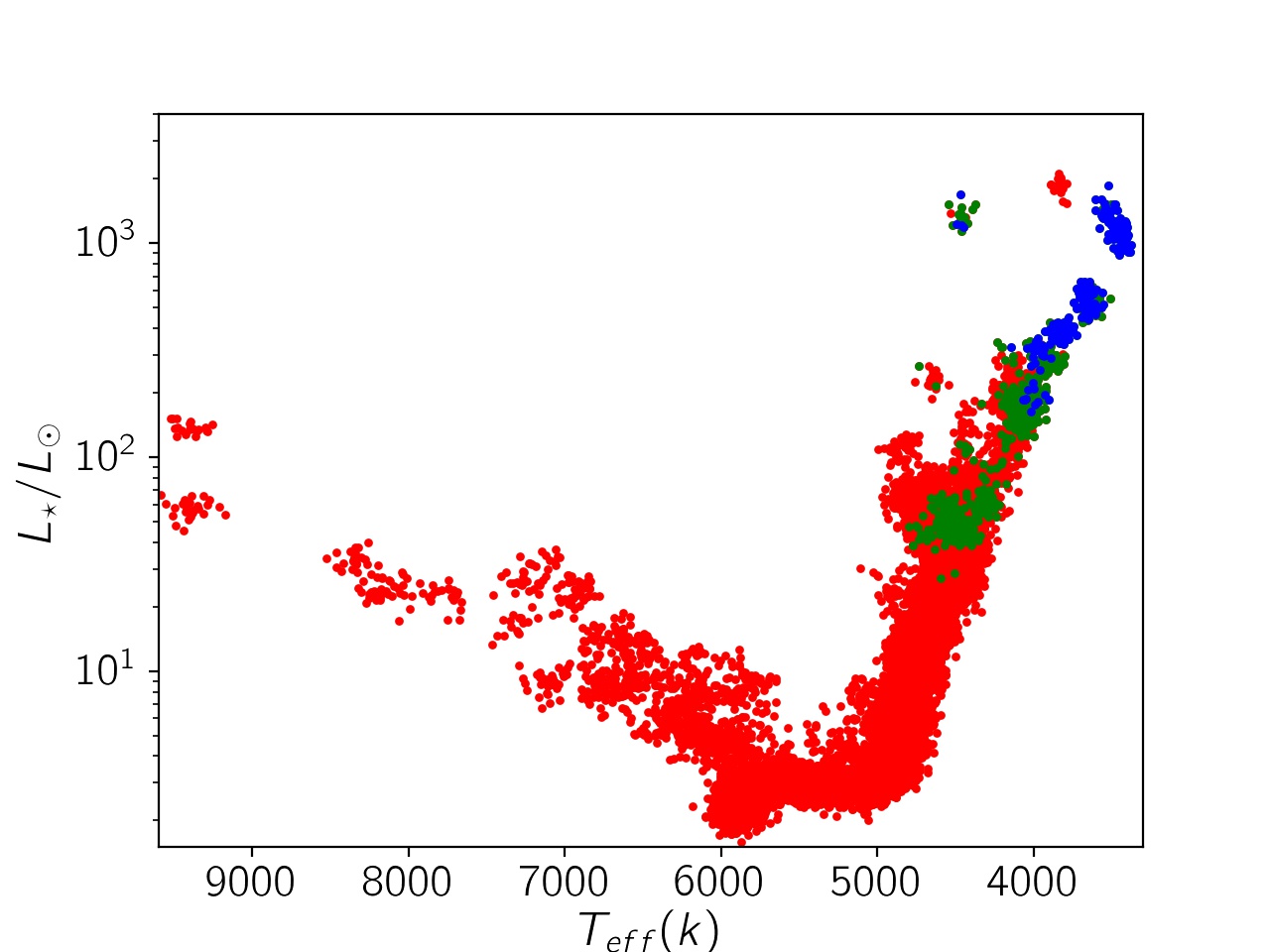}
	\caption{Left panel: The histogram of the scattering optical depth (in the logarithmic scale) for a large sample of cool RGB stars. Right panel:  Hertzsprung-Russell diagram of simulated cool RGB stars with the red dots. The green filled dots represent the giant stars with the scattering optical depth larger than $0.01$. The blue dots show the source stars whose polarization signals can be detected in microlensing events. }\label{optical}
\end{figure*}

\subsection{Red Clump Giants}
According to the H-R diagram, giant stars evolve from red giant branch $\rm{(RGB)}$ to asymptotic giant branch $\rm{(AGB)}$ over time \citep[see, e.g.,][]{Carroll}. The light variation of AGB stars in terms of time changes in the range of a few tens of days to more than $1000~\rm{days}$ \citep{lebzelter2011}. From the Besan\c{c}on model \citep{robin2003,robin2012} for simulating the stellar populations, we do not include AGB stars as the source stars of the microlensing event as the majority of AGB stars at the Galactic bulge vary with a period of $P < 13~\rm{days}$. 
Traditionally, all the variables in the microlensing studies are considered as the background events and excluded from the analysis. In the exceptional studies as \citep{2006OGLE}, they investigate the microlensing of variable stars. These stars compose the population of only $1/10$ of stars with constant baseline. Based on this strategy, we also exclude AGB stars in our analysis.

The hot giant stars do not compose a large fraction of RGB stars compared to the cool giant stars in the Galactic bulge \citep{Nucita}. In what follows, we study only the cool RGBs as the source of microlensing events. The cool giant stars have extended circumstellar atmospheres with dust grains that produce effective polarization of light. We follow the formalism developed by \citet{simmons2002} to investigate the polarization signal by the red giant stars with a significant amount of dust in the atmosphere. Figure (\ref{simon}) shows  schematic representation of the cross-section of a cool RGB star with the radius of $R_\star$, a cavity, and an envelope.   
The dust grain in the extended atmosphere of cool RGB star forms at the temperature of $1400~\rm{K}$, the so-called inner circumstellar radius of the atmosphere is $R_{\rm{h}}$ and the outer radius is $R_{env}$. Here, we adapt the assumption of  $R_{env} \gg R_{\rm{h}}$. The inner radius is a function of the stellar temperature and is given by  {\citep[see,e.g.,][]{lamers, Ingrosso2015}:  
\begin{eqnarray}
R_{h}=0.45(\frac{T_{eff}}{T_{h}})^{2.5} R_{\star},
\label{star}
\end{eqnarray} 
where $T_{\rm h}\simeq 1400~\rm{K}$ is the temperature that the condensation of dust occurs.
The dust formation occurs above the condensation radius ($R_h$) and observation of polarized signal could determine $R_{\rm{h}}$ and the dependency of this parameter to other parameters of the stellar atmosphere. The cooler giant stars have smaller $R_{\rm{h}}$ and larger polarization signal. For the cool giant stars with the effective temperatures in the range of $3000-6000~\rm{K}$, the inner circumstellar radius is in the range of $3-17~R_{\star}$.

\noindent The density of the circumstellar envelope is given by \citep{simmons2002}
\begin{equation}
n(r)=
\left \{ \begin{array}{c}
0  \quad ~~~~~~~~~ (r<R_{h})\\
n_{0}(\frac{R_{h}}{r})^{\beta} \quad (r \geq R_{h} )
\end{array} 
\right \}
\end{equation}
where we adapt $\beta=2$ \citep{simmons2002} and $n_{0}$ is the dust number density at the inner radius. The polarized Stokes intensity linearly depends on the scattering optical depth "$\tau_{\rm sc}$" where the scattering optical depth is proportional to the scattering cross-section $\sigma_{\rm sc}$ and the surface number density of dust (integrating over $z$-axis from the density distribution).  The relation between the polarization signal and optical depth parameters is given in Appendix C.
The scattering optical depth is a function of the mass-loss rate, terminal dust velocity, and $R_{h}$ \citep{Ignace2008}. Accordingly, the scattering optical depth is given by \citep{Ignace2008}:  
\begin{equation}
\tau_{sc}= 2\times 10^{-3} \eta \kappa (\frac{\dot{M}}{10^{-9}M_{\odot} yr^{-1}})(\frac{30 kms^{-1}}{v_{\infty}})(\frac{24R_{\odot}}{R_{h}}),
\label{tau1}
\end{equation}
 This equation could derive from the equations ($B1$ to $B5$) in Appendix B. where $\eta$ is the dust to the gas mass density ratio, $v_{\infty}$ is the terminal dusty wind velocity and $\kappa\simeq 200$  $cm^{2} g^{-1} $ is the the dust opacity for $\lambda>5500$ $\mathring{A}$.
In this regard, \citet{Origlia2002} derived the wind velocity of cool RGB stars in the Galactic globular clusters using Spectral Energy Distribution (SED ) modeling as a function of the stellar luminosity:
\begin{eqnarray}
v_{\infty}= 14  (\frac{L_{\star}}{1000 L_{\odot}})^{0.3} (\frac{\varrho}{200})^{-0.5}   (km/s),
\label{vinfty}
\end{eqnarray}
The equation ($11$) is normalized to $14$ $km/s$ for luminosity $L_{\star}=1000 L_{\odot}$ and $\varrho$ =$200$ of a stellar source.
where $\varrho= 1/\eta =200 \times 10^{-(Z+0.76)}$ is gas to dust ratio and $Z$ is the stellar metallicity and $L_\star$ is the luminosity of the source star. 
\begin{figure*}
	\centering
	\includegraphics[angle=0,width=0.49\textwidth,clip=]{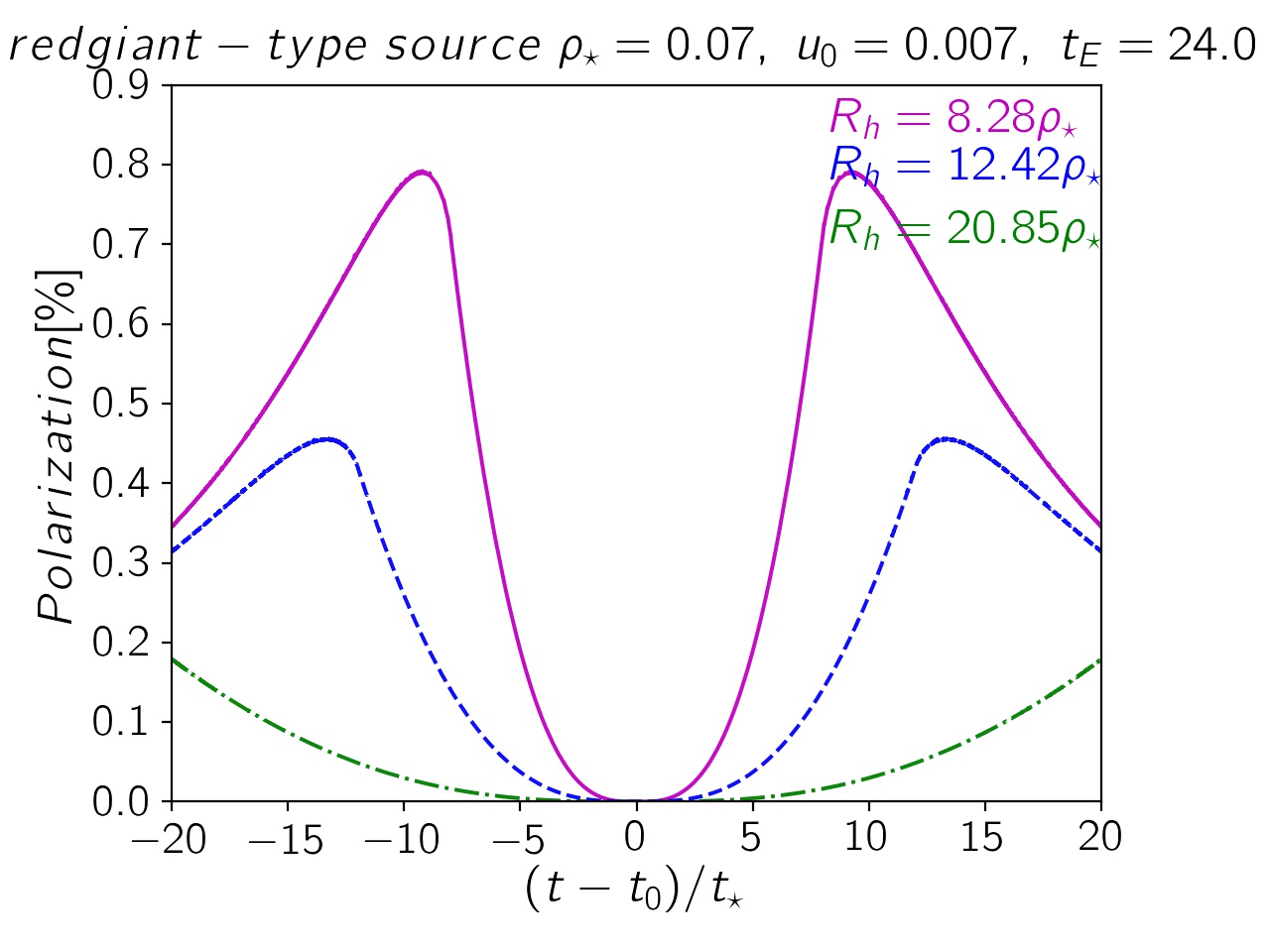}
	\includegraphics[angle=0,width=0.49\textwidth,clip=]{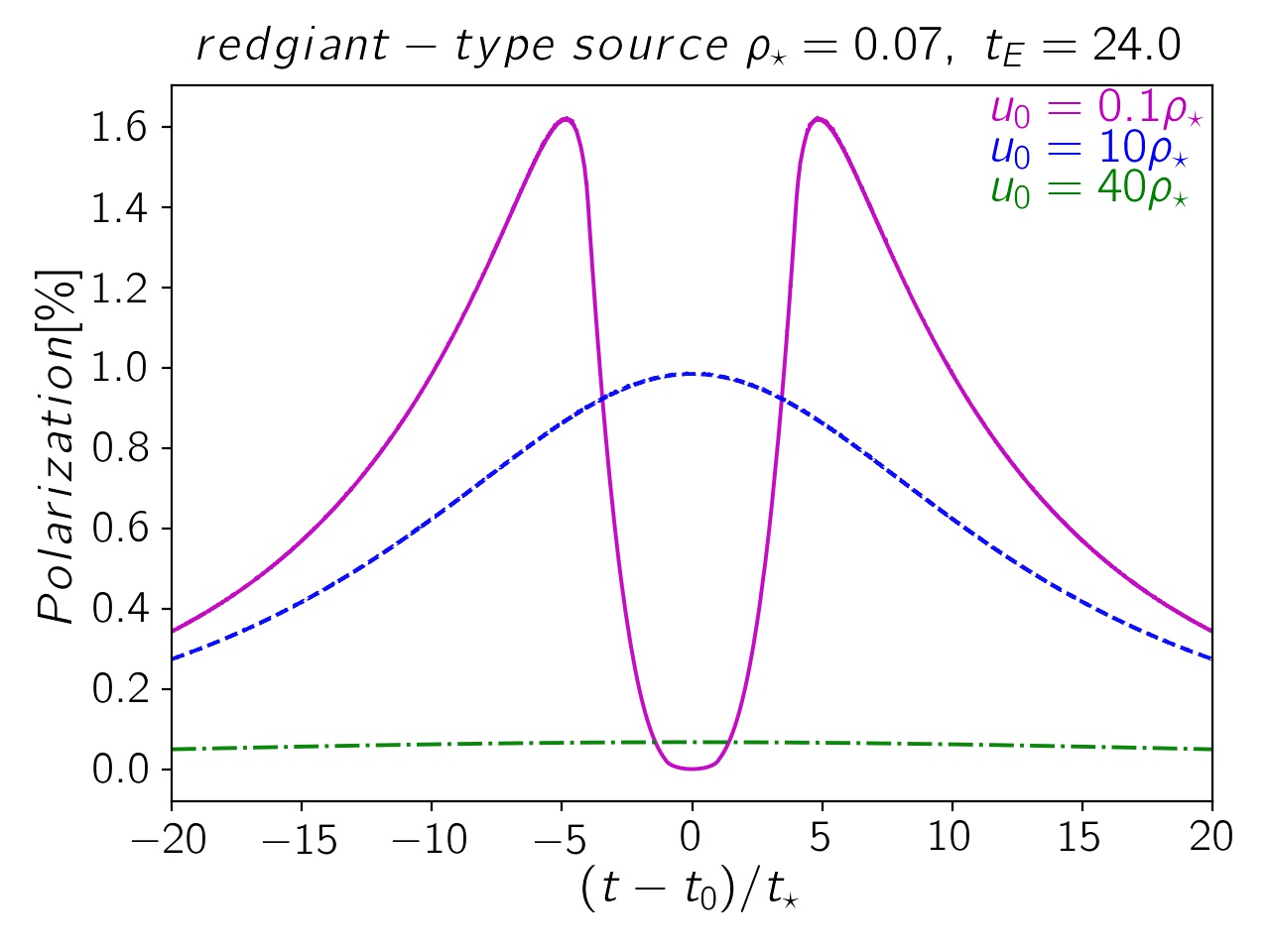}
	\caption{Left panel: the polarization curves of a lensed M$7$ super-giant star in a transit microlensing event. The purple, blue and green curves are due to different amounts of $R_{h}$ which have resulted in the scattering optical depth amounts $0.08$, $0.05$, and $0.03$, respectively. Right panel: Similar to the previous plot, but for three different amounts of the lens impact parameter.}\label{superg}
\end{figure*}

\noindent To calculate the mass loss of stars in the cool RGB phase, $\dot{M}$, we use the empirical formula given by \citet{Reimers}: 
\begin{equation}
\dot{M}= 4\xi \times 10^{-13}\frac{({L_\star}/{L_{\odot}})}{({g_{\star}}/{g_{\odot}})({R_{\star}}/{R_{\odot}})}(M_{\odot} yr^{-1}),
\label{mdot}
\end{equation}
that depends on the luminosity, the stellar radius and the surface gravity of the source star. This relation has a constant of proportionality ($ \xi$) that we use the value of  \citet{Ingrosso2015} for cool RGB stars in the bulge galaxy, i.e., $ \xi \simeq3.0$.

\noindent Substituting Equation (\ref{vinfty}) and (\ref{mdot}) in Equation (\ref{tau1}), the scattering optical depth obtain as:
\begin{eqnarray}\label{tau}
\tau_{sc} =15\times 10^{-5} \frac{\kappa ({L_\star}/{L_{\odot}})^{0.7}}{({M_\star}/{M_{\odot}})~({T_{eff}}/{T_{h}})^{2.5}~\varrho^{0.5}}.
\end{eqnarray}
 Hence, in the cool RGB stars, the brighter and cooler giants with higher metallicity have higher scattering optical depths and these stars are suitable candidates for the polarimetry follow-up observations of the microlensing events.

 We generate a population of cool RGB stars in the Galactic bulge with the Besan\c{c}on model \citep{robin2003,robin2012}. The left panel of Figure (\ref{optical}) shows the distribution of scattering optical depth for cool RGB stars toward the Galactic bulge. A small sample of the giant stars has a considerable  scattering optical depth ($\tau_{\rm{sc}}>0.01$). These stars in the right panel in the  Hertzsprung-Russell diagram of cool RGB stars are shown with the green circles. From this figure,  the scattering optical depth of the $\rm{M}$ and $K$-type giant stars are considerably larger than the other cool RGB stars. These stars are $1000$ times brighter than a sun-like star and their effective surface temperature is relatively low at around $3500~\rm{K}$.

 In Figure (\ref{superg}), we show the polarization curve of the M$7$ supergiant star in the microlensing events with different sizes of the inner circumstellar radius. We use the codes developed by \citet{martin} for the extended source microlensing events for calculating magnification and polarization factor. We use GNU Scientific Library to calculate the complete elliptical integrals in the code
 \footnote{http://www.gnu.org/software/gsl/index.html}.
 The scattering optical depth as well as the polarization signal decrease with increasing the size of the inner circumstellar radius of the extended atmosphere of cool RGB stars.
 We use the transit time for the 
 polarization detection where the lens passes across the extended atmosphere of the source star as \citep{Ingrosso2015}
 \begin{eqnarray}
 \delta t=2 t_{E}\sqrt{(\frac{R_{h}}{0.75})^{2}-u_{0}^{2}},
 \end{eqnarray}
 and with measuring this time scale we can determine the size of the extended atmosphere of the source star.
 At the right panel of Figure (\ref{superg}),  we show the effect of the lens impact parameters on the polarization curves. Increasing the lens impact parameter decreases the polarization signal significantly.

\section{Monte-Carlo simulation}\label{three}

 In this section, we produce synthetic microlensing events towards the Galactic bulge. By taking into account the atmospheric models for the source stars, we measure the polarization of events during microlensing. We assume that microlensing events are monitored by the OGLE and MOA surveys and the follow-up telescopes. We use the specifications of the FORS2 polarimetry on VLT in this simulation.

\begin{figure}
 	\begin{center}
 		\includegraphics[angle=0,width=0.49\textwidth,clip=]{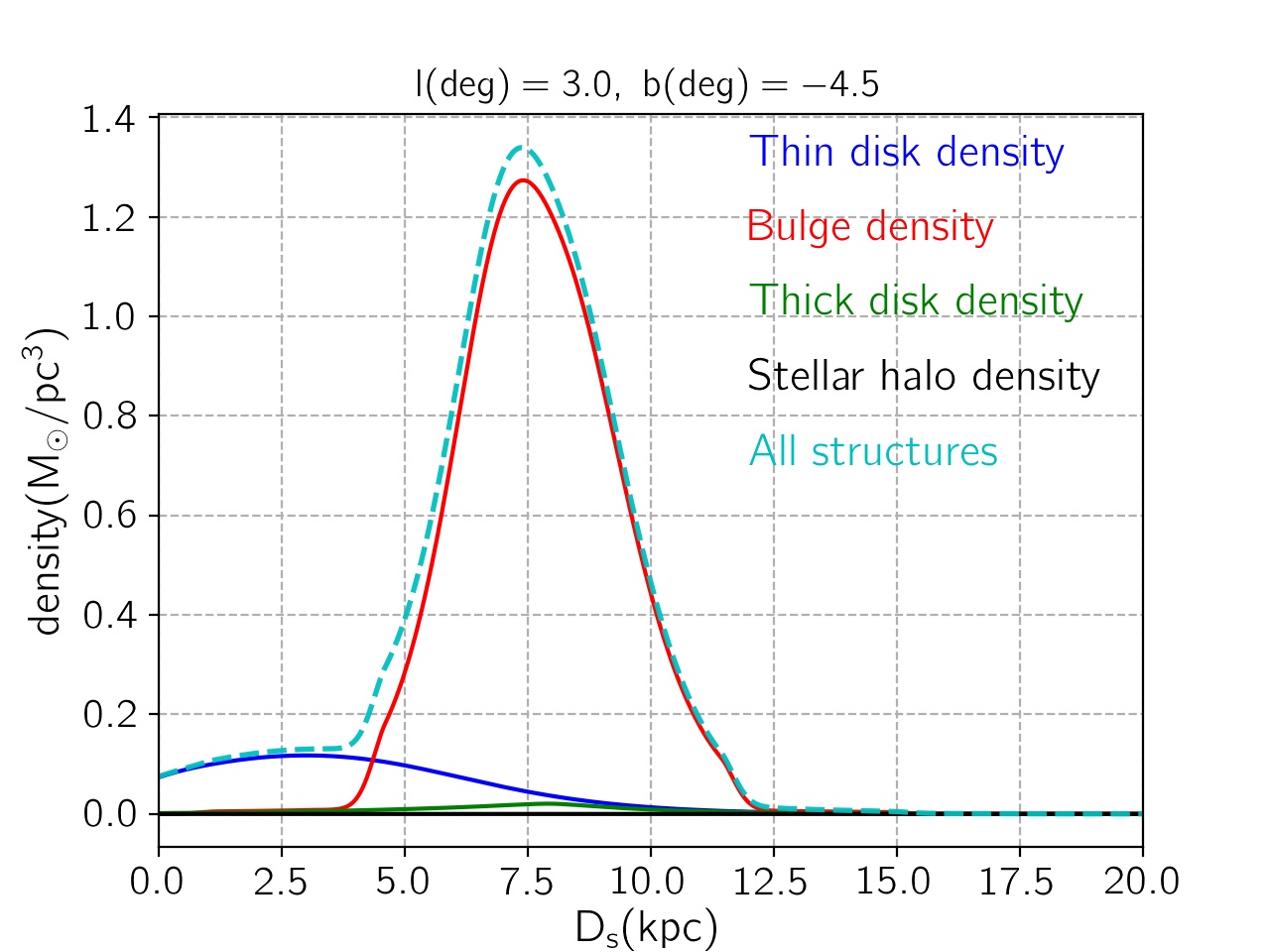}
 		\caption{In the plot, the mass densities of the Galactic bulge (red solid curve), thin disk (blue solid), thick disk (green solid) and stellar halo (black solid) versus the distance from the observer toward $(l,b)=(3.0,-4.5)~$deg are shown. The overall density is represented by the dashed cyan curve.}
 		\label{sourced}
 	\end{center}
 \end{figure}
\subsection{Simulation of microlensing events}
 To simulate the microlensing events, we need to generate the source star and lens populations toward the Galactic bulge. The simulation is done toward the Galactic bulge fields with $l\in [-3:3]~\rm{deg}$ and $b\in[-5.5:-3.5]~\rm{deg}$ which coincides with the several OGLE fields.\\ 
For producing the source star population, after fixing a line of sight, i.e., $(l, b)$ values, we choose the source distance from the observer ($D_{\rm s}$) using the probability, i.e., $dP/dD_{\rm s}(l, b) \propto \rho_{\rm t}(l, b, D_{\rm s}) D_{\rm s}^{2}$.  Here, $\rho_{t}= \rho_{\rm b}+ \rho_{\rm d} +\rho_{\rm h}$ is the overall mass density due to the Galactic bulge, (thin and thick) disks and the stellar halo in the given line of sight $(l, b)$. These distributions are taken from the Besan\c{c}on model \citep{robin2003,robin2012}. As an example for the Galactic mass densities along the direction of $(l, b)=(3.0, -4.5)$deg is shown in Figure (\ref{sourced}) with the contribution of each structure in the overall mass density.
For each structure in the galaxy, we make a big ensemble of its stars by indicating their photometry properties, mass, metallicity, age, surface gravity, stellar type, and their luminosity class using the new version of the Besan\c{c}on  model \footnote{\url{https://model.obs-besancon.fr/}}.  These ensembles help to specify the source properties in each galactic structure. The Besan\c{c}on model simulates the stellar distributions and their characterizations in the bulge, discs, and halo of our galaxy based on the Padova Isochrone \citep{padova}.

In the next step, we indicate the source apparent magnitude by adding the distance module and extinction to its absolute magnitude. The extinction values are determined using the 3D extinction map given by \citet{Marshal2006}. They measured 3D $K_{s}-$extinction map for the extensive ranges of the Galactic longitude and latitude with the step $\Delta b=\Delta l= 0.25^{\rm{\circ}}$. Then, the extinction in $K_{s}-$band is converted to its value in $K-$band using $A_K=0.95~A_{K_s}$ \citep{Marshal2006}. The $K-$band extinction is in turn converted to the extinction in other bands using the relations given by \citet{Cardelli1989}. We assume $R_V$ of $2.5$ for Galactic bulge and $3.1$ for thin disk, thick disk, and the stellar halo \citep{Nataf2013,Cardelli1989}. Finally, a Gaussian fluctuation is added to each extinction value with the width in the range $[0.017, 0.04]$ depending on the wavelength \citep{Cardelli1989}. The map of the extinction in $I-$band for the simulated fields with $l\in[-3, 3]$deg and $b\in[-5.5, -3.5]$deg is represented in Figure \ref{extinc}. Accordingly, the extinction value changes in the range $[0.5, 1.6]$ for these fields and its average value is $1.1~$mag.

The lens population is generated using the Besan\c{c}on model. Their distance are chosen from the microlensing rate function, i.e., $\Gamma \propto \rho_{t}\sqrt{D_{\rm l}(1-\frac{D_{\rm l}}{D_{\rm s}})}$. The complimentary details of the Monte-Carlo simulation can be found in \citet{moniez,sajadian2019}.

 \begin{figure*}
 	\begin{center}
 		\includegraphics[angle=0,width=0.95\textwidth,clip=]{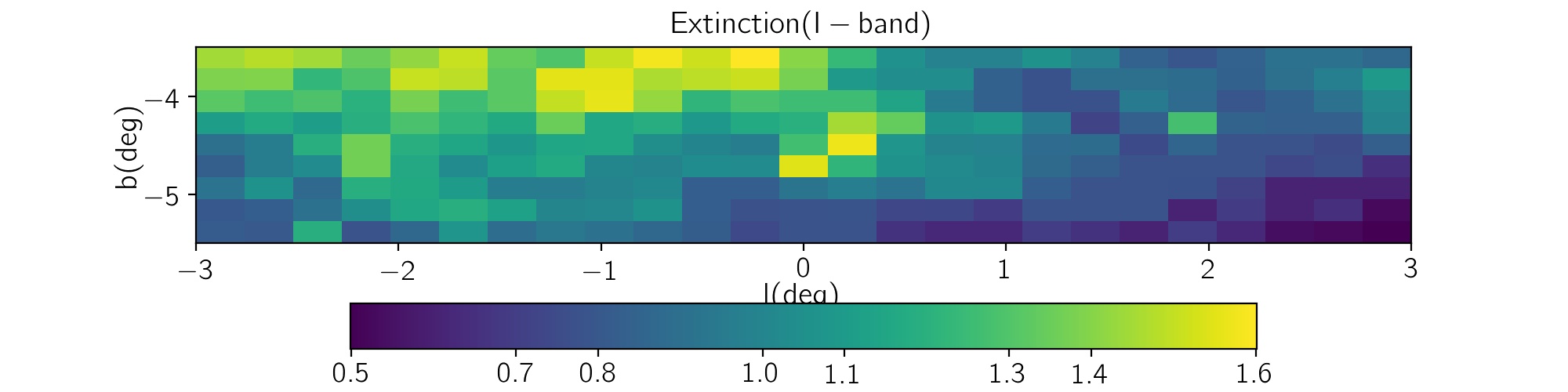}
 		\caption{The map of $I-$band extinction toward the Galactic bulge fields with $l\in[-3, 3]$deg and $b\in[-5.5, -3.5]$deg.}
 		\label{extinc}
 	\end{center}
 \end{figure*}

We also select the microlensing events based on the detection efficiency of OGLE survey which depends on the Einstein crossing time, i.e., $\epsilon(t_E)$ \citep{Wyrzykowski2015b}. 
We ignore (i) the events with $t_{E}$ either larger than $300~$d or less than $0.5$ d, (ii) the events whose source stars are either white dwarf, T-Tauri, AGB stars, (iii) the events with source stars brighter than $12$ or fainter than $21$ magnitude in $I$-band. We adapt the apparent magnitude of the source stars within the range of $12$ $<$ $I$ $<$ $21$ as expressed in the OGLE catalog \citep{2019OGLE}.
and (iv) the high-blending events.

\begin{figure*}
	\begin{center}
		\includegraphics[angle=0,width=0.49\textwidth,clip=]{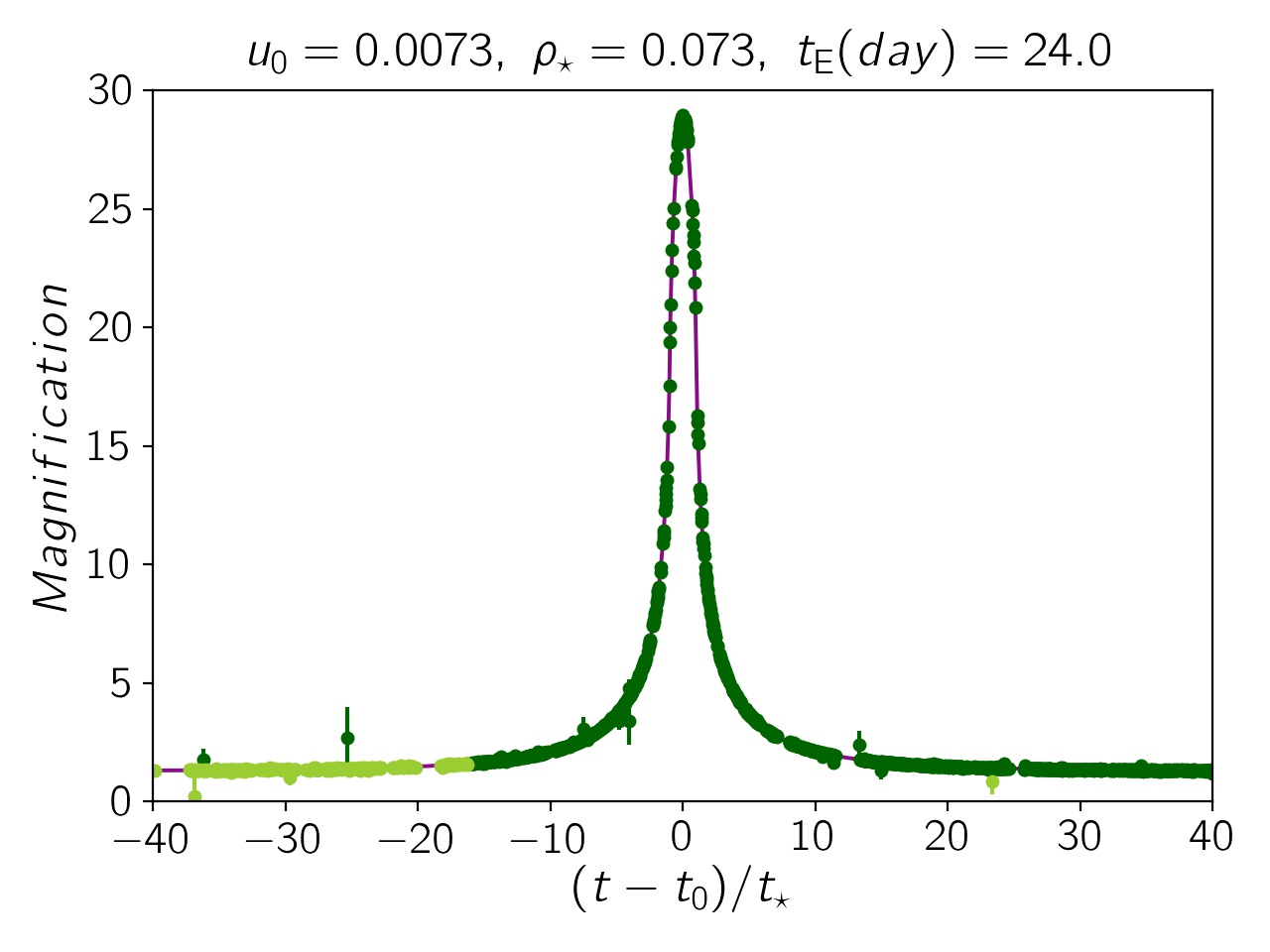}
		\includegraphics[angle=0,width=0.49\textwidth,clip=]{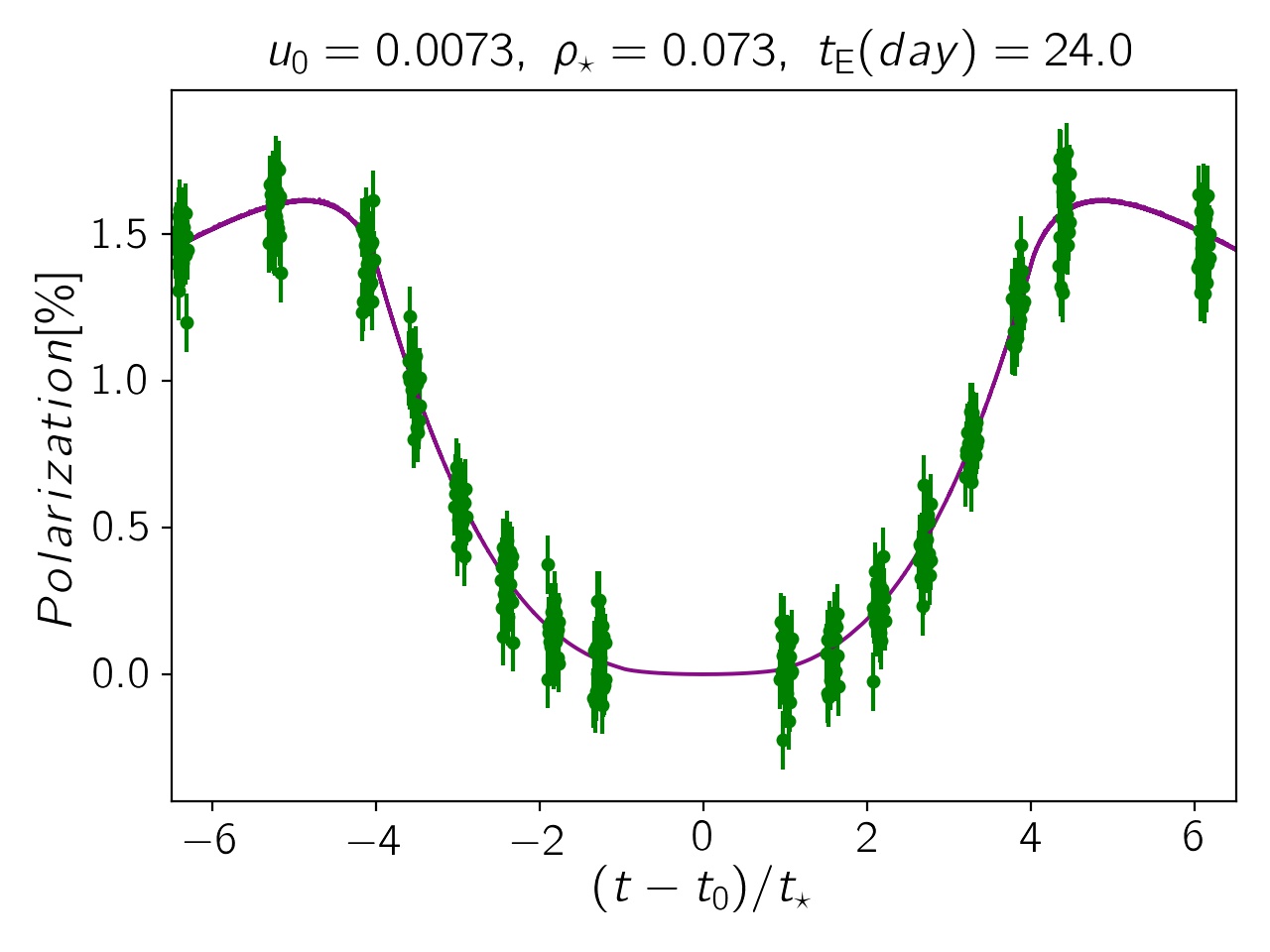}
		\caption{ In the left panel: a sample of simulated microlensing event for a red giant source star. The data taken by survey and follow-up groups are shown with light and dark green, respectively. The right panel represents the polarimetry curve. The parameters used in this simulation are given at the top of the plots. }
		\label{lightsimss}
	\end{center}
\end{figure*}

\subsection{Simulation of polarimetric and photometric data}\label{four}
\textbf{Photometry}: In the next step, we produce synthetic data points for modeling microlensing events. We use the archived data points of the gravitational microlensing events from the survey and follow-up telescopes \citep{sajadian2016} to have a realistic sampling rate and the photometric precision for the light curves. We assume photometry observations being done in the range of $[-3 t_{\rm E}:3 t_{\rm E}]$, setting $t_0$ in the middle of this range, also $t_{0}$ is uniformly chosen in the range of $[0,9]~\rm{months}$. We note that the microlensing photometry observations are performed by the survey telescopes such as OGLE, MOA, and KMTNet towards the center of the Galactic bulge. They monitor millions of stars in the bulge direction to find out any variations in the light curve of stars. There is a live-automatic alert system to identify the microlensing events within the millions of stars. When the microlensing event is identified and the magnification reaches to a threshold, e.g., ${3}/\sqrt{5}$, survey telescopes release the coordinates of the source star as the microlensing candidate (i.e. Right ascension(RA) and Declination ascension (DA)) as well as $t_0$ and $t_{\rm E}$ of the ongoing event to the follow-up telescopes.

To simulate the light curve of each event,  the data points of the light curves are shifted by a Gaussian probability function with the width of the photometric error respect to the theoretical light curve. The left panel of Figure (\ref{lightsimss}) shows an example of simulation from the microlensing event with the source star of a supergiant. In this simulation, we produce data points both from the survey and follow-up telescopes.

 \textbf{Polarimetry}: For simulating the polarimetry curves, we use the theoretical polarizations on the source stars and generate the corresponding microlensing light curves. We assume that the observation is performed by the VLT telescope, using FORS2 polarimetry. We note that there are other polarimetries such as ZIMPOL, PEPSI. However, the disadvantage of these polarimeters is that they are working only for the bright stars (i.e $<12$) which is not suitable for the microlensing targets. Here, we perform our simulation for FORS2 polarimetry in VLT.

 We assume the start time of the polarization observations randomly is chosen in the range of $[-7t_{\star}: 7t_{\star}]$. The observation period is taken to be six hours per night. Each data point needs 18 min of the observation. This time is composed of the exposure time plus the readout time. The observation can be interrupted due to the bad weather and possibly other targets of opportunity observations. In our simulation, we put the constraint of a lower bound on the measured polarimetry to be larger than $0.1$ percent. This is the optimum value of the precision for FORS2 polarimeter. For simulating the polarimetry light curves,  the polarization at each point is shifted with a Gaussian function according to the error of the polarimeter \citep{sajadian2016b}.
 
In our Monte-Carlo simulation, we choose those events that at least $3$ consecutive polarimetry data points to be above the baseline with the distances of $1\sigma$ to  $4\sigma$.
 Regarding the criteria that we adapted for detection, the main-sequence stars are not feasible for polarization detection. For the stars out of main-sequence stars, we exclude the variable stars, white dwarfs (as they are faint), and hot RGB stars (low population). We limit our observations to cool RGB stars where due to dust formation in their extended atmosphere, the polarization signal is considerable.

In the caption of Table (\ref{tab2}), we define the characteristics of polarimetry microlensing detection of events with cool RGB source stars such as: microlensing parameters ($\log_{10}[t_{\rm{E}}]$, $\rho_{\star}$, and $u_{0}$) and metallicity. Other columns of this table show the atmosphere parameters ($\tau$, $R_{h}$, and surface gravity), type, and luminosity class of the stars
	\footnote{\url{https://model.obs-besancon.fr/modele_help.html}}. The last $4$ columns identify the flag of detection criteria with $1\sigma$ to $4\sigma$ signals. 
\section{Results and discussions}\label{five}
\subsection{Statistics}
We simulate a big ensemble of microlensing events that are detectable by the OGLE survey. The detectability criteria are (i) the apparent source magnitude to be in the range of $[12, 21]~$mag, and (ii) passing the detection efficiency versus the Einstein crossing time \citep[Fig (8) of ][]{Wyrzykowski2015b}. 
Two scatter-histogram plots of the simulated microlensing events are shown in Figure (\ref{giantplot}): the scattering optical depth versus the stellar types of giant sources (left panel) and the lens impact parameter versus the Einstein crossing time (right panel). The events with detectable polarization signals are shown with light blue filled circles and columns. Accordingly, most of the observable events are due to $K$ and $M$~type giant source stars with relatively larger scattering optical depth. The cool and inflated bright giant source stars in microlensing events are the best candidate for microlensing events for polarimetry follow up observations.  In the H-R diagram, these source stars are at the top and right corner (see right panel of Figure \ref{optical}). 
Our simulations reveal that the polarization signals for the cool RGB source stars with the scattering optical depth (i.e. $\tau_{sc}$) less than $0.01$ can not be discerned by FORS2.

\begin{figure*}
		\includegraphics[angle=0,width=0.49\textwidth,clip=]{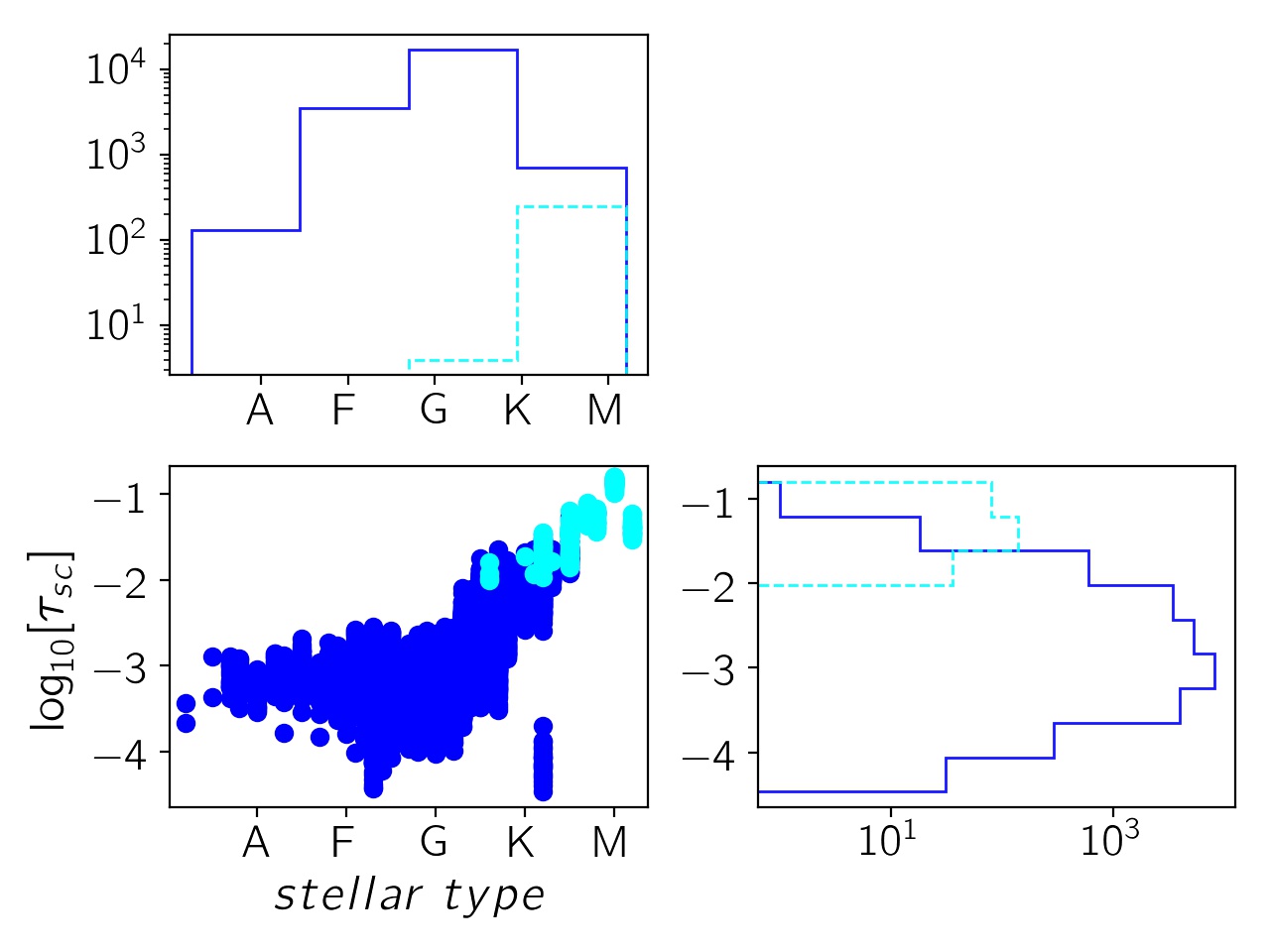}
		\includegraphics[angle=0,width=0.49\textwidth,clip=]{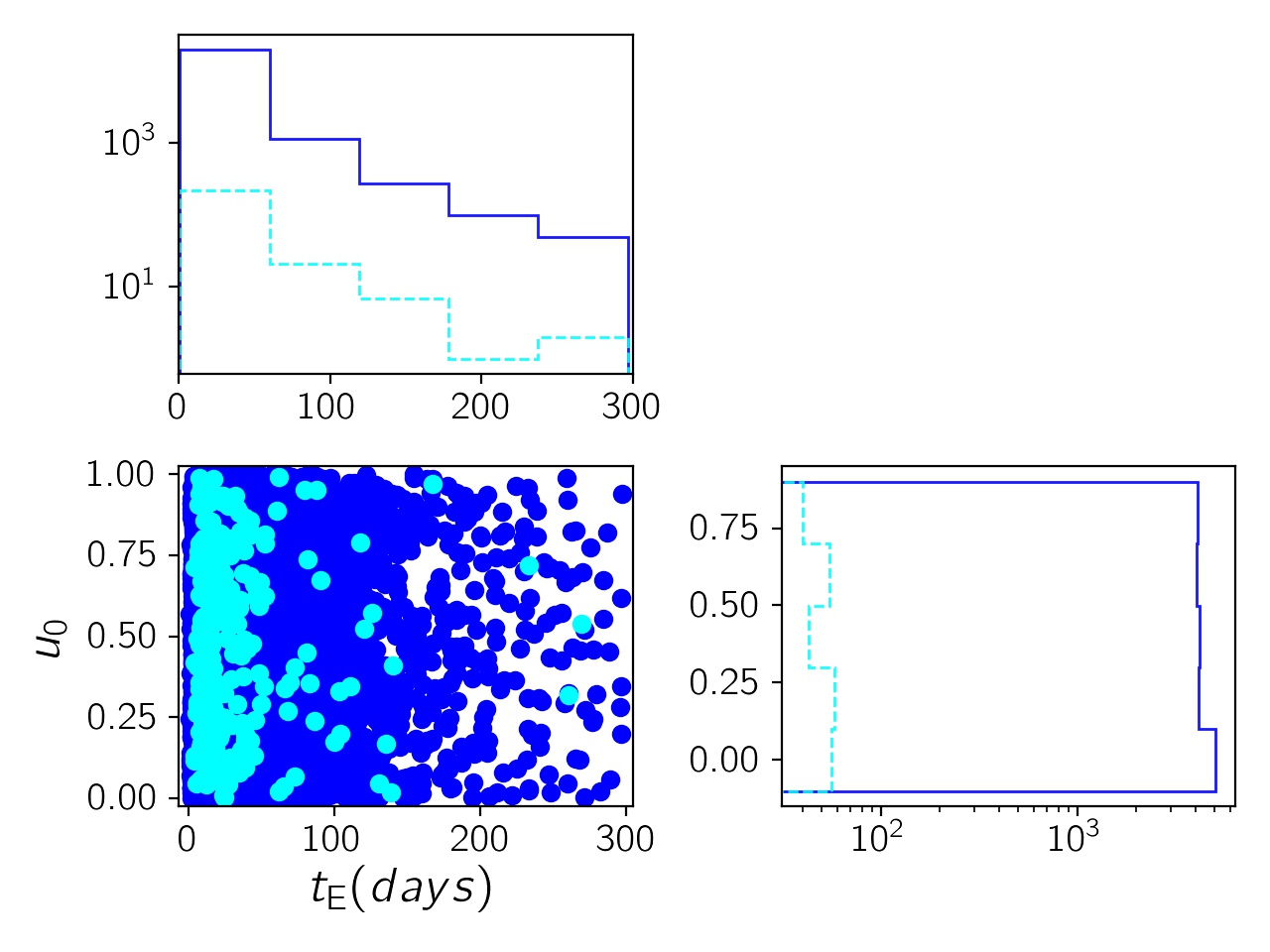}
		\caption{ Left panel: the scatter-histogram plot of the scattering optical depth of the giant source stars versus the stellar type for all simulated microlensing events. Right Panel: the scatter-histogram plot of the lens impact parameter versus the Einstein crossing time of the simulated microlensing events.  In both plots, the detected events are indicated with light blue filled circles and columns.}
		\label{giantplot}
\end{figure*}

The number of microlensing events with detectable polarization signals (i.e. $N_{ml,p}$) in the observations can be given by:  
\begin{equation}
N_{ml,p}= \frac{\Delta N_{ml}}{\Delta t} \times T_{obs}\times \epsilon_p 
\end{equation}
 where $\Delta N_{ml}/\Delta t$ is the number of microlensing events detected by OGLE, KMTNet and MOA surveys per year which is about $\sim 3000$/yr \citep{OGLEIII, KMTNet} and $ \epsilon_p$ is the fraction of observed microlensing events with detectable polarization signal. 
For one year of observation (i.e. $T_{\textrm{obs}}=1~$yr) towards the Galactic bulge with the simulation of the OGLE survey, we obtain that the expected number of  microlensing events with detectable polarization signals are $N_{\rm{ml,p}}\sim 20,~10,~8, $ and $5$ for the four different criteria of being three consecutive data points above the baseline with $1\sigma$, $2\sigma$, $3\sigma$,  and $4\sigma$, respectively.

\subsection{Error estimations for physical parameters}\label{six}
In this section,  we use the results of simulations for each event and recover the parameters of the lens and source as well as the errors associated to these parameters.
We use the covariance matrix method that is the inversion of the Fisher matrix to calculate the errors of parameters. 
We note that for a given light curve, there might be degenerate solutions. Here we do not investigate these solutions.
The Fisher matrix is defined as \citep[see, e.g.,][]{Verde2007,Heavens}
\begin{eqnarray}
F_{i,j}=\langle \frac{\partial^{2} \ln L }{\partial \theta_{i}  \partial \theta_{j}}\rangle,
\end{eqnarray}
where $L$ is the likelihood function. $ \theta_{i}$ and $ \theta_{j}$ are the free parameters of the model and brackets represents the averaging. The error propagation on the parameters is computed from the matrix $\sqrt{F^{-1}}$ \citep[see, e.g.,][]{Belokurov}. Diagonal and off-diagonal elements of this matrix show the uncertainties of parameters and their correlations, respectively. The Fisher matrix of photometry and polarimetry observations can be given as: 
\\
\begin{equation}
F_{i,j}=\sum_{k=1}^{N} \sigma_{P}^{-2} \frac{\partial P(t_{k})}{\partial \theta_{i}}  \frac{\partial P(t_{k})}{\partial \theta_{j}} +  \sigma_{A}^{-2} \frac{\partial A(t_{k})}{\partial \theta_{i}}  \frac{\partial A(t_{k})}{\partial \theta_{j}}. 
\end{equation}
\\
 Here $A$ and $P$ are the magnification factor and the polarization signal, respectively. $\sigma_{A}$ and $\sigma_{P}$ are the photometry and polarimetry uncertainties. There are four parameters which we would like to determine them with the corresponding errors from the simulated photometry and polarimetry data points, namely: $t_{\rm{E}}$, $u_{0}$, $\rho_{\star}$, and $\tau$.\\
 As shown in the right panel of Figure (\ref{superg}) and the paper \citep{simmons2002}, the size of cavity could be calculated by the time-width between the polarization peaks,  
so polarimetry observation could determine the inner radius $R_{h}$ of the source star. In our simulation, we use equation (\ref{star}) for calculation of $R_{h}$.  This parameter is a function of source star radius and temperature which are can be determined with spectroscopic observations or the color-magnitude diagram. 
  Due to low polarimetry precision, we combine the polarimetry and photometry data to explore the parameters of the source star atmosphere.  We derive the parameters from fitting the photometry light curves and photometry $+$ polarimetry light curves based on Fisher and covariance matrix analysis. Inlcuding the polarimetry observations not only increases the accuracy of parameters related to the photometry observations but also enables us to constrain the physical parameters of the source star as the optical depth $\tau$ and the extended radius of the source star (Table (\ref{tab4})).  
We note that the polarimetry is a function of scattering optical depth (i.e. $\tau$) of the atmosphere where it depends on the physical parameters of the source star as defined in equation (\ref{tau}). 
In this equation, we could determine $L_\star$ and $T_{eff}$ with the apparent magnitude measurement and color-magnitude diagram respectively.  Additional parameters as the metallicity of the atmosphere (i.e. Z) can also be determined from spectroscopic observations and the mass of source star could be identified regarding the stellar evolution and the position of star in color-magnitude diagram.

Using the combination of the polarimetry and photometry data with the knowledge of the source-star parameters, we can determine the dust opacity of the atmosphere of cool RGB source stars and the extended radius of the stars where dust is formed.
As shown in table (\ref{tab4}), the polarimetry data increases the accuracy in measurements of microlensing parameters. We note that the polarimetry data has large error bars and using only this data, we can not  measure accurately the microlensing parameters. The photometric observations can provide three parameters
of microlensing events as $t_{E}$, $u_{0}$, and $\rho_{\star}$ where $\rho_{\star}$ is 
measurable in the high magnification microlensing events. 
The advantage of polarimetry 
observations is that, it not only measures the parameters related to the source stars also improve the accuracy of parameters that we determine from the photometry observations. The results of table (\ref{tab4}) are presented in Figure (\ref{diss}) where the combined photometry and polarimetry data is 
compared with the photometry data. Note that the microlensing parameters such as $\rho_{\star}$ for some events have large error bars: so we ignore them and represent a sub-sample of events from table (\ref{tab4}).

\begin{figure*}
	\begin{center}
		\includegraphics[angle=0,width=0.49\textwidth,clip=]{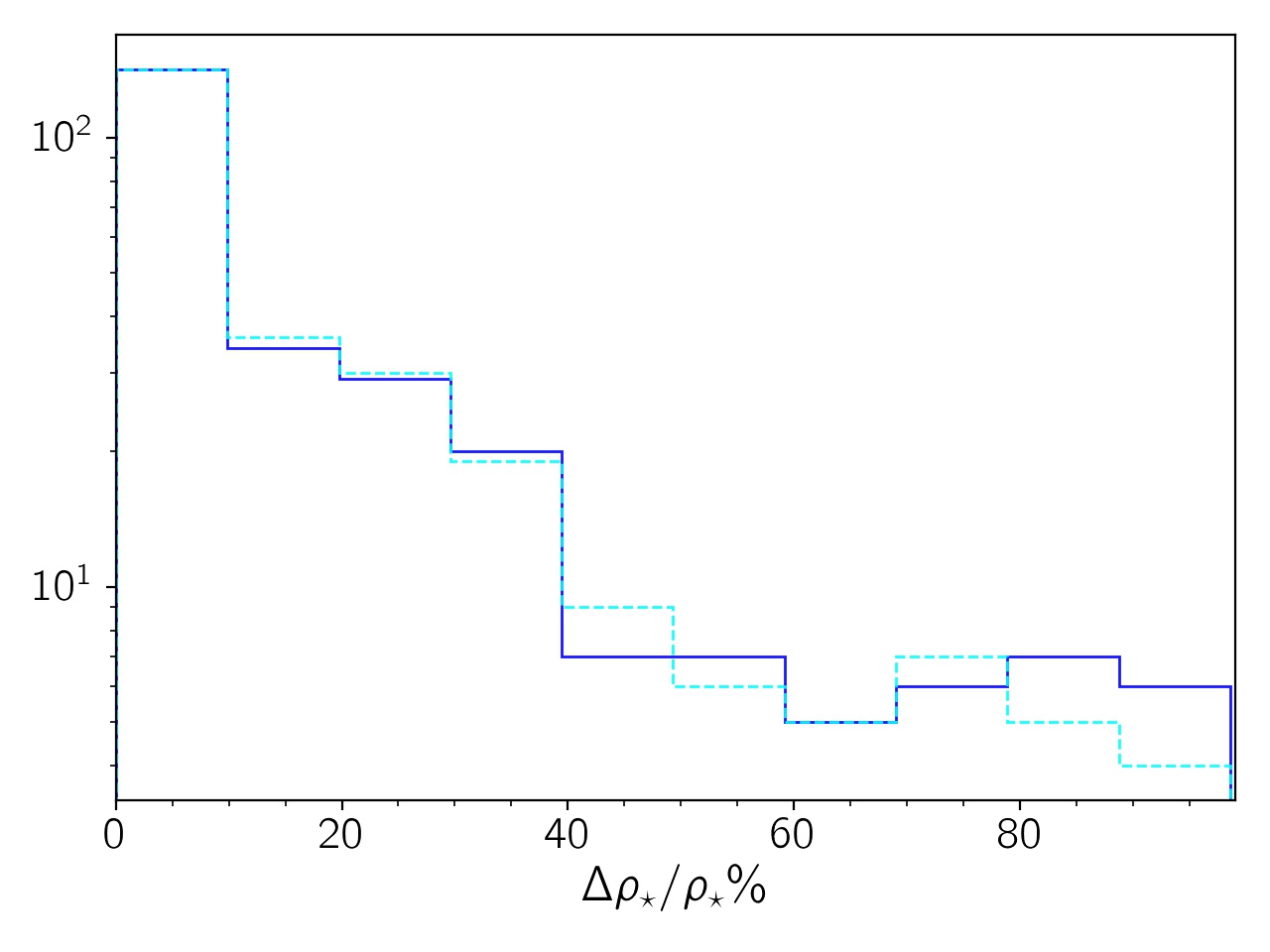}
		\includegraphics[angle=0,width=0.49\textwidth,clip=]{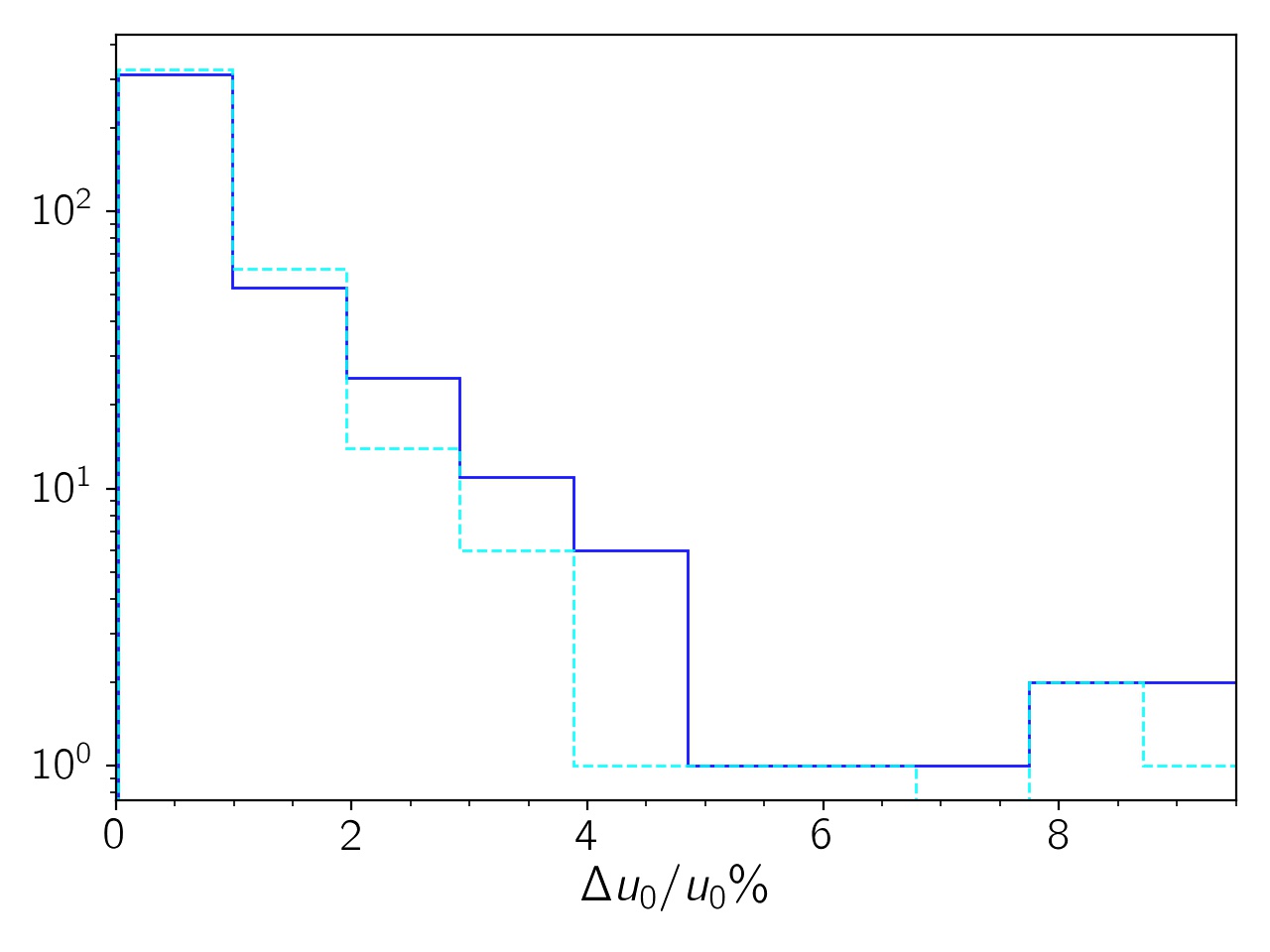}
		\includegraphics[angle=0,width=0.49\textwidth,clip=]{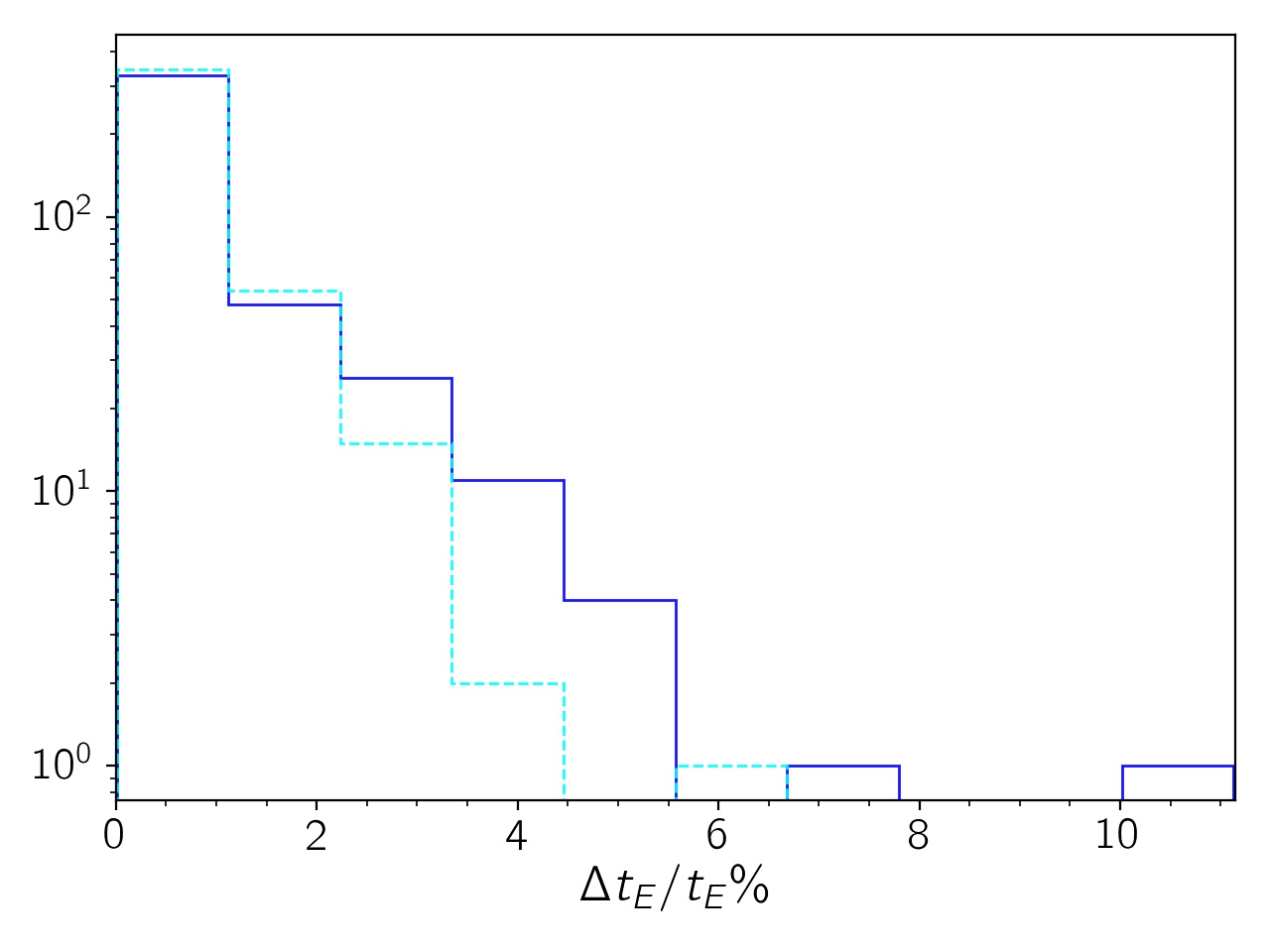}
		\includegraphics[angle=0,width=0.49\textwidth,clip=]{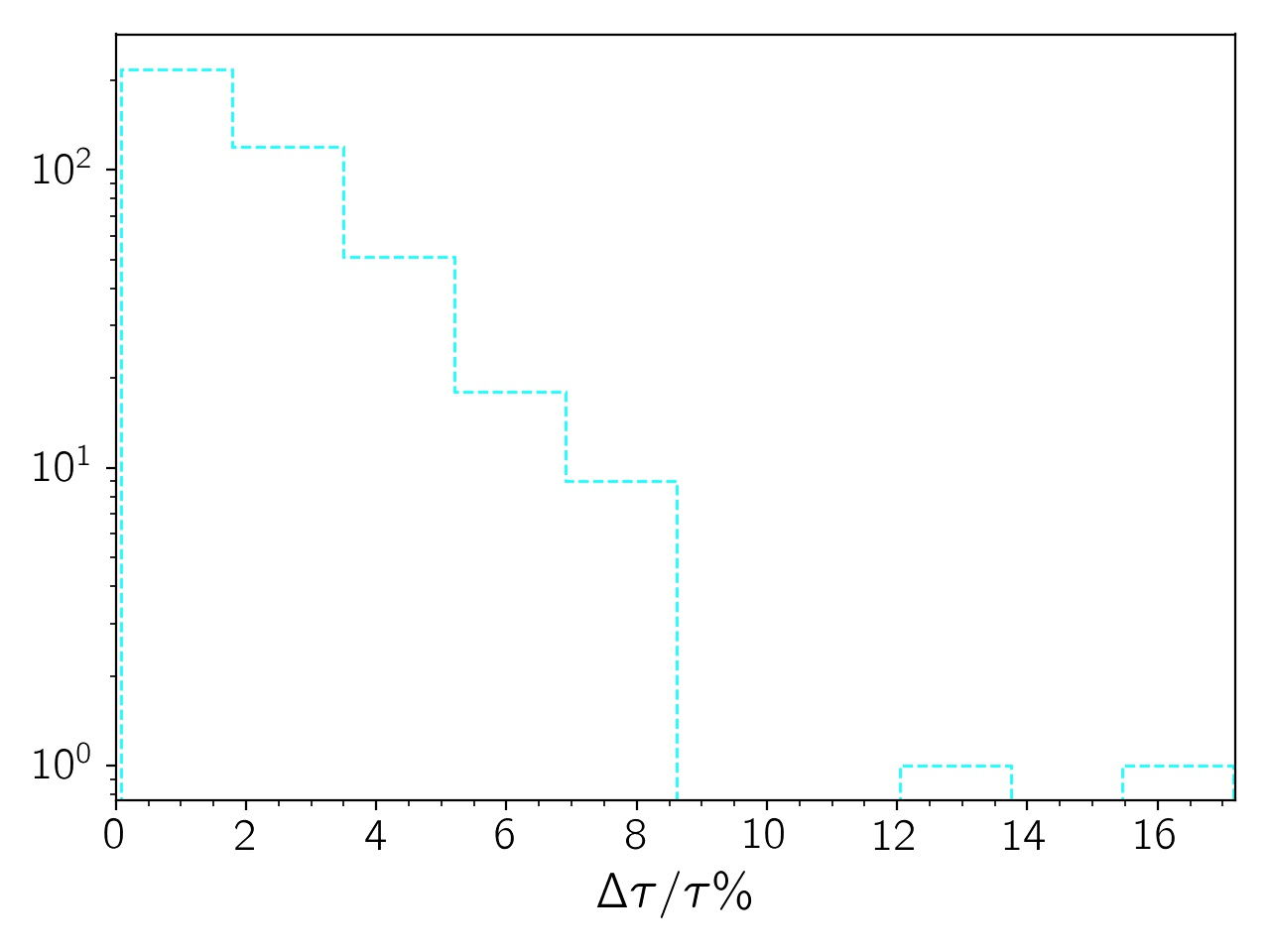}
		\caption{ Distribution of the relative error of the cool RGB source star parameters ($t_{E}$, $u_{0}$, and $\rho_{\star}$) with photometry (blue- solid line) and polarimetry$+$photometry (light blue-dash line) observations. Also, the distribution of scattering optical depth parameter is determined with polarimetry+photometry observation. }
		\label{diss}
	\end{center}
\end{figure*}

\section{Conclusions}
\label{summary}
  In this work, we investigated the possibility of detection of polarization signals from a source star during the gravitational microlensing. In this regard, we first extended the formalism for calculating the polarization signal in microlensing events and studied the polarimetry signal of main-sequence stars from the O-type stars to the M-type stars in various filters with different stellar surface gravity and their surface temperature. In the shorter wavelengths, the polarization signals are stronger. Also, the early-type stars with lower surface gravities have more polarization signals in microlensing events.
  
  We also studied the polarization signals of the red giant source stars. In this type of stars, the polarization is proportional to the scattering optical depth, which is formulated in Equation (\ref{tau}) (more details are given in Appendix B and C). Accordingly,  the brighter and cooler giant stars ($M$ and $K-$type ones) with higher metallicity have higher scattering optical depth and as a result high polarization signals in microlensing events.

  To evaluate the detectability of polarization signals in microlensing events, we performed a Monte Carlo simulation for the main sequence and the red giant source stars. The probability of detecting the polarization signal for the main-sequence stars with VLT-FORS2 is small. However, our simulations anticipated that the probability of detecting polarization signals of red giant source stars are $20,~10,~8 $, and $5$  per year for the four criteria of being three consecutive data points above the baseline with $1-4$ $\sigma$ where we assumed detection of about  $3000$ microlensing events during one year with the survey telescopes. 
  We also examined the improvement of parameter measurement of microlensing events with the polarimetry observations besides the photometry observations. We used the covariance matrix analysis to evaluate the error estimation of parameters. To conclude, the polarimetry observations decrease the errors in the determination of the microlensing parameters ($t_{E}$, $u_{0}$, and $\rho_{\star}$). These parameters are constrained with photometry observation. Also, the polarimetry observation could determine other parameters of the stellar atmosphere (i.e. scattering optical depth and inner radius
  of the stellar envelope of cool RGB stars).  These two parameters contain information about the dust opacity and the radius where dust is formed.
 
\section*{Acknowledgements}
We thank J. P. Harrington who kindly provided for us the polarization profiles of early-type stars and R. Ignace for helpful comments and his useful notes on the manuscript. We are grateful to V. Bozza for insightful comments and discussions. S. Sajadian thanks the Department of Physics, Chungbuk National University and especially C. Han for hospitality. 
This research was supported by Sharif University of Technology Office of Vice President for Research under Grant No. G950214. Numerical computations were performed on the cluster of
Department of Physics, University of Salerno. We would like to thank anonymous referee for his/her useful comments. 

\section*{DATA AVAILABILITY}
The data underlying this article are available in the article and in its online supplementary material.

\bibliographystyle{mnras}
\bibliography{references1}

\begin{table*}
	\centering
	\caption{Characteristics of the microlensing events with the detectable polarization signals due to the cool RGB source stars. The columns are (i) the number of events, (ii) the logarithm of Einstein crossing time, (iii) the source angular size normalized to the Einstein angle, (iv) the minimum impact parameter of the lens trajectory, (v) the metallicity of source, (vi) the scattering optical depth of source atmosphere, (vii) the surface gravity of the star, (viii) the inner circumstellar radius of extended circumstellar atmosphere normalized to $R_{E}$ and projected on the lens plane, (ix) the luminosity class of source star, and (x) the type of source star (for more information about $c_{l}$ and type of star, see \text{\url{https://model.obs-besancon.fr/modele_help.html}}). 
	The other parameters are the detectability criterion of event that shows with $1$ if at least 3 consecutive polarimetry data points are greater than of $1 \sigma$, $2 \sigma$, $3 \sigma$, and $4\sigma$ respectively.
	    A complete electronic version of this table is available online.}
	\label{tab2}
	\begin{tabular}{cccccccccccccc}
		\hline \hline
		{\tiny }	
		$\rm No.$	& $\log_{10}[t_{\rm{E}}]$ & $\rho_{\star}$ & $u_{0}$ & $Z(Z_{\odot})$ & $\tau$ & $\log_{10}[g]$ & $R_{h}$ & $c_{l}$ & $type$ & $1 \sigma$ & $ 2 \sigma$ & $3 \sigma$ & $ 4 \sigma $ \\
		&$\rm{(day)}$&& & & &   $\rm{(cm~s^{-2})}$ &&&&&&& \\ \hline	
		$ 1 $ & $ 1.71 $ & $ 0.28 $ & $ 0.35 $ & $ 2.30 $ & $ 0.05 $ & $ 0.88 $ & $ 1.29 $ & $ 3 $ & $ 7.3 $ & $ 1 $ & $ 1 $ & $ 1 $ & $ 1 $ \\
		$ 2 $ & $ 1.62 $ & $ 0.17 $ & $ 0.69 $ & $ 1.32 $ & $ 0.14 $ & $ 0.28 $ & $ 0.74 $ & $ 3 $ & $ 7.5 $ & $ 1 $ & $ 1 $ & $ 1 $ & $ 1 $ \\
		$ 3 $ & $ 1.91 $ & $ 0.10 $ & $ 0.45 $ & $ 1.32 $ & $ 0.14 $ & $ 0.28 $ & $ 0.45 $ & $ 3 $ & $ 7.5 $ &  $ 1 $ & $ 1 $ & $ 1 $ & $ 1 $ \\
		$ 4 $ & $ 1.59 $ & $ 0.27 $ & $ 0.10 $ & $ 1.32 $ & $ 0.14 $ & $ 0.28 $ & $ 1.14 $ & $ 3 $ & $ 7.5 $ & $ 1 $ & $ 1 $ & $ 1 $ & $ 1 $ \\
		
		\\ \\ \hline 
	\end{tabular}
\end{table*}	
\begin{table*}
	\centering
	\caption{The relative error of the cool RGB source star parameters with photometry and polarimetry$+$photometry observations. A complete list is available online.}
	\label{tab4}
	\begin{tabular}{ccccccc}
		\hline \hline
		$\rm No.$	&$\frac{\Delta t_{\rm{E}}}{t_{\rm{E}}}\%$ &$\frac{\Delta \rho_{\star}}{\rho_{\star}}$$\%$ & $\frac{\Delta u_{0}}{u_{0}}$$\%$ & $\frac{\Delta \tau}{\tau}$$\%$  \\ \hline
		$ 1 $ & $ 2.45 $ & $ 55.73 $ & $ 12.37 $ & $ .... $  \\
		& $ 1.73 $ & $ 37.50 $ & $ 8.33 $ & $ 13.63 $  \\ \\
		
		$ 2 $ & $ 0.33 $ & $ 8.17 $ & $ 0.37 $ & $ .... $  \\
		& $ 0.31 $ & $ 7.69 $ & $ 0.35 $ & $ 0.20 $  \\ \\
		
		$ 3 $ & $ 0.69 $ & $ 22.56 $ & $ 0.87 $ & $ .... $ \\ 
		& $ 0.48 $ & $ 14.30 $ & $ 0.56 $ & $ 0.80 $  \\ \\
		
		$ 4 $ & $ 0.08 $ & $ 0.06 $ & $ 0.57 $  & $ .... $  \\
		& $ 0.08 $ & $ 0.06 $ & $ 0.57 $ & $ 0.41 $ \\

	\end{tabular}
\end{table*}	
\appendix
\section{Limb-darkening and polarized coefficients}
In Table (\ref{tab1}), we mention the limb-darkening and polarized coefficients for different types of stars and different filters $UBVRI$. These parameters are inferred from fitting polynomial functions of $1-\mu$ to the Stokes (the total and polarized) intensities that are estimated in \citep{Harrington15}.\\
\begin{table*}
\centering
\caption{The columns are the number of events, effective temperature,  logarithm of surface gravity, filters (The $0$, $1$, $2$, $3$, and $4$ correspond to filters $UBVRI$ respectively), and the Stokes intensity coefficients $a_{i}$ and $b_{i}$ for different stellar types. Here we assume the stellar metallicity to be almost equal to the sun metallicity. A complete electronic version of this table is available online.}
\label{tab1}
\begin{tabular}{ccccccccc}
\hline \hline
&&&$\rm No.$&$T_{eff}$& $\log_{10}[g]$ & $Filter$ &&\\ 
&&&& $(K)$&  $\rm{(cm~s^{-2})}$ &&&\\  
$a_{0}$ & $a_{1}$ &$a_{2}$&$a_{3}$& $a_{4}$ &$a_{5}$&$a_{6}$& $a_{7}$ &$a_{8}$\\
$b_{0}$ & $b_{1}$ &$b_{2}$&$b_{3}$& $b_{4}$ &$b_{5}$&$b_{6}$& $b_{7}$ &$b_{8}$\\ \hline
& & & $ 0 $ & $ 15000 $ & $ 2.0 $ & $ 0 $ & & \\ 
$ 0.0424 $ & $ -0.0146 $ & $ -0.0784 $ & $ 0.5976 $ & $ -2.4293 $ & $ 5.4071 $ & $ -6.7150 $ &  $ 4.3596 $ & $ -1.1539 $ \\ 
$ -0.0000 $ & $ 0.0019 $ & $ -0.0383 $ & $ 0.3088 $ & $ -1.2641 $ & $ 2.8408 $ & $ -3.5604 $ & $ 2.3331 $ & $ -0.6229 $ \\ \\

& & & $ 1 $ & $ 15000 $ & $ 2.0 $ & $ 1 $  &&\\ 
$ 0.0310 $ & $ -0.0115 $ & $ -0.0232 $ & $ 0.1589 $ & $ -0.6573 $ & $ 1.4781 $ & $ -1.8596 $ &  $ 1.2249 $ & $ -0.3302 $ \\ 
$ -0.0000 $ & $ 0.0010 $ & $ -0.0185 $ & $ 0.1489 $ & $ -0.6123 $ & $ 1.3819 $ & $ -1.7400 $ & $ 1.1459 $ & $ -0.3077 $ \\ \\

& & & $ 2 $ & $ 15000 $ & $ 2.0 $ & $ 2 $ &&\\ 
$ 0.0142 $ & $ -0.0024 $ & $ -0.0481 $ & $ 0.3791 $ & $ -1.5559 $ & $ 3.5010 $ & $ -4.3968 $ &  $ 2.8882 $ & $ -0.7739 $ \\ 
$ -0.0000 $ & $ 0.0007 $ & $ -0.0118 $ & $ 0.0956 $ & $ -0.3922 $ & $ 0.8836 $ & $ -1.1104 $ & $ 0.7297 $ & $ -0.1955 $ \\ \\

& & & $ 3 $ & $ 15000 $ & $ 2.0 $ & $ 3 $ &&\\
 $ 0.0085 $ & $ 0.0001 $ & $ -0.0530 $ & $ 0.4247 $ & $ -1.7386 $ & $ 3.9061 $ & $ -4.8959 $ & $  3.2089 $ & $ -0.8574 $ \\ 
 $ -0.0000 $ & $ 0.0005 $ & $ -0.0082 $ & $ 0.0665 $ & $ -0.2726 $ & $ 0.6133 $ & $ -0.7695 $ & $ 0.5049 $ & $ -0.1350 $ \\ \\
 
& & & $ 4 $ & $ 15000 $ & $ 2.0 $ & $ 4 $ \\ 
$ 0.0046 $ & $ 0.0013 $ & $ -0.0500 $ & $ 0.4030 $ & $ -1.6462 $ & $ 3.6922 $ & $ -4.6183 $ &  $ 3.0200 $ & $ -0.8046 $ \\ 
$ -0.0000 $ & $ 0.0003 $ & $ -0.0051 $ & $ 0.0411 $ & $ -0.1679 $ & $ 0.3771 $ & $ -0.4724 $ & $ 0.3094 $ & $ -0.0825 $ \\
\\ \hline
\end{tabular}
\end{table*}

\section{the Optical depth calculation}
 The polarimetry data is mainly a function of optical depth along the radial direction of the star. This parameter can be taken as a global parameter of the atmosphere of the source star:
\begin{equation}
\tau= \int_{R_{h}}^{\infty} n \sigma dx= \frac{1}{m_{d}}\int\rho_{d} \sigma dx=\frac{1}{m_{d}}\int \eta\rho_{g} \sigma dx
\end{equation}
Where $\rho_{d}$ is the dust density and $\rho_{g}=n_{0}(\frac{R_{h}}{r})^{\beta}m_{A}$ is the mass density of gas. $m_{d}$ and $m_{A}$ are the dust mass and the atomic mass respectively.  $\sigma$ is the scattering cross section and $n$ is the scatterer number density (Simmons et al. $2002$). The optical depth is
\begin{equation}
\tau=\eta (\frac{m_{A}}{m_{d}}) \sigma \frac{n_{0} R_{h}}{\beta-1}=\frac{\eta}{m_{d}}\sigma \frac{\rho_{gas} R_{h}}{\beta-1}
\end{equation}
As we know $ \dot{M}=4 \pi R_{h}^{2} \rho_{g}  v_{\infty} /g_{v}$
\begin{equation}
\tau= \frac{\eta}{m_{d}}\sigma \frac{\dot{M}}{(\beta-1)4 \pi  (v_{\infty}/g_{v}) R_{h} }
\end{equation}
Here $g_{v}=2$ allows for the formulation of wind velocity law \cite{Ignace2008}. Therefore the $\tau$ is expressed as:	
\begin{equation}
\tau= \eta \kappa \frac{\dot{M}}{2 \pi v_{\infty} R_{h} }
\end{equation}
By using equation $(12)$, the scattering optical depth parameter obtains as:
\begin{equation}
\tau_{sc} =15\times 10^{-5} \frac{\kappa ({L_\star}/{L_{\odot}})^{0.7}}{({M_\star}/{M_{\odot}})~({T_{eff}}/{T_{h}})^{2.5}~\varrho^{0.5}}.
\end{equation}

\section{the Stokes intensities of cool RGB stars}
The Stokes flux for cool RGB stars normalized to $F_{\star}$ (the Flux of the source star) is as follows:
\begin{eqnarray}
 \frac{F}{F_{\ast}}=
\left ( \begin{array}{c}\\
H_{\ast}(pl)+\tau_{sc}H_{I}(pl)\\
\tau_{sc}H_{p}(pl)\cos(2 \phi)\\
- \tau_{sc}H_{p}(pl)\sin(2 \phi)\\
0 
\end{array}\right).
\end{eqnarray}
Where $H_{\ast}(pl)$, $H_{I}(pl)$, and $H_{p}(pl)$ are given by $C4$, $C5$, and $C6$ equations respectively.
$R_{h}$ is the inner circumstellar radius of the atmosphere of cool RGB star and  $\tau_{sc}$ is the scattering optical depth. $A(\rho,\alpha)$ is the magnification factor.
The integration is taken over the source surface with the projected radius $\rho_{\star}$ normalized to the Einstein radius. $\rho$ is the radial component over the surface of the star which is normalized to $\rho_{\star}$ and $\phi$ is the azimuthal angle over the source surface.
Each element of the surface source star are determined by polar coordinate $(\rho, \alpha)$. $pl$ shows the distance between the lens and the center of source star.
So the magnification and polarization are given by:
\begin{equation}
 M(pl)=H_{\ast}(pl)+\tau_{sc}H_{I}(pl)
\end{equation}
\begin{equation}
 P(pl)=\frac{\tau_{sc}H_{p}(pl)}{H_{\ast}(pl)+\tau_{sc}H_{I}(pl)}
\end{equation}
\begin{equation}
H_{\ast}(pl)= \frac{1}{\pi \rho_{\star}^{2}}\int_{0}^{2\pi} d\alpha \int_{0}^{\rho_{\star}} A(\rho,\alpha) \rho d\rho
\end{equation} 
\begin{equation}
H_{I}(pl)=\frac{3(\beta-1)}{16 \pi} R_{h}^{\beta-1}\int_{0}^{2\pi} d\alpha \int_{0}^{\infty} A(\rho,\alpha) \Theta_{I}(\rho) d\rho
\end{equation}  
\begin{equation}
H_{p}(pl)=\frac{3(\beta-1)}{16 \pi} R_{h}^{\beta-1}\int_{0}^{2\pi}  \cos(2 \alpha) d\alpha \int_{0}^{\infty} A(\rho,\alpha) \Theta_{p}(\rho) d\rho
\end{equation}
Here $\Theta_{I}(\rho)$ and $\Theta_{p}(\rho)$ are the functions of $\rho$ that are introduced in equations $25$ and $26$ of \citet{simmons2002}.

\label{lastpage}
\end{document}